\documentclass{article}
\usepackage[T1,T2A]{fontenc}
\usepackage[utf8]{inputenc}
\usepackage[russian,ukrainian,english]{babel}
\usepackage{latexsym,amssymb,amsmath}
\usepackage{graphicx}
\usepackage{color}

\def\vcntr#1{\raisebox{-0.5\height}{#1}}

\def\GeV{{\rm\ GeV}}
\def\ve{\varepsilon}
\def\CG{{\cal G}}
\def\CM{{\cal M}}
\def\be{\begin{equation}}
\def\ee{\end{equation}}
\def\bea{\begin{eqnarray}}
\def\eea{\end{eqnarray}}
\def\Re{\mathop{\rm Re}\nolimits}
\def\Im{\mathop{\rm Im}\nolimits}
\def\tg{\mathop{\rm tg}\nolimits}

\def\<{\langle}
\def\>{\rangle}

\def\para{\parallel}
\def\up{\uparrow}
\def\down{\downarrow}

\def\half{{^1\!/_2}}

\def\Jou#1#2#3#4#5#6{{\it #1}. #2 {\bf #3}, #4 (#5).}
\def\ea{{\it et al}}

\def\la{\lambda}

\begin{document}

\title{Two photon exchange in elastic electron scattering off hadronic systems}
\author{Dmitry~Borisyuk$^1$, Alexander~Kobushkin$^{1,2}$\\[2mm]
\it\small $^1$Bogolyubov Institute for Theoretical Physics, 14-B Metrologicheskaya street, Kiev 03680, Ukraine\\
\it\small $^2$National Technical University of Ukraine "Igor Sikorsky KPI", 37 Prospect Peremogy, Kiev 03056, Ukraine}
\maketitle

\begin{abstract}
In the present review we discuss different aspects of the two-photon exchange (TPE) physics in elastic $ep$ scattering, at high $Q^2$ as well as at low $Q^2$.
 
The imaginary part of the TPE amplitude gives rise to beam and target single-spin asymmetries. Different theoretical approaches to calculation of these observables are considered.
The real part of the TPE amplitude influence unpolarized cross section and double-spin observables and is, most likely, responsible for discrepancy between two methods of proton form factors measurements.

We review different methods of calculations of the TPE amplitudes the framework of ``hadron'' and ``quark-gluon'' approaches. We discuss the dispersion approach suitable for low and intermediate $Q^2$, which includes elastic and inelastic intermediate hadronic states, as well as connection of TPE to proton radius puzzle.
 
The present situation with direct experimental searches for the TPE amplitude in $e^+p/e^-p$ charge asymmetry is also discussed, as well as attempts to extract the TPE amplitudes from existing experimental data obtained by the Rosenbluth and double polarization techniques.

The TPE physics in other processes, such as elastic $\mu p$, $e$-nucleus and $e\pi$ scattering is also reviewed.
\end{abstract}

\tableofcontents
%%%==================================================================================
\section{Introduction}
Understanding of the internal structure of the proton, neutron, and other strongly interacting systems was for a long time one of the fundamental problems of particle physics. Experiments on elastic and inelastic scattering of ultra-relativistic electrons off nucleons and nuclei provide unique tool for such a study. 

Early experiments with relativistic electron beams scattering off hadron systems were done under the leadership of Robert Hofstadter in 1950's at High Energy Physics Laboratory (HEPL) at Stanford \cite{Hof1} and gave information on the radii of wide spectrum of nuclei, as well as the distribution of electric charge in them. 

Later on, the same method was used to measure the proton size. The proton radius of 0.77 fm, extracted in 1955 from the cross sections of elastic electron-proton scattering with the electron beams of energy up to 550 MeV \cite{Hof2}, testified irrefutably that the proton is not ``elementary'' particle and has internal structure. A lot of interesting information on this stage of experiments at HEPL is contained in the review paper by Hofstadter~\cite{Hof3}.

After that, a lot of experiments were done to measure the proton size more accurately. The present measurements give for the proton radius 0.8775(51)~fm from the electron-proton scattering \cite{CODATA} and 0.84087(39)~fm from atomic transitions in the muonic hydrogen \cite{rE-muon-0,rE-muon}. The reason of difference between these data is not yet clear and is discussed intensively in literature
(very recently, the results of PRad experiment were published, where the value of 0.831(7)(12)~fm was obtained from $ep$ scattering at very low $Q^2$ \cite{PRad}).

The first information on the internal magnetic structure of the neutron was reported in 1958 \cite{Hof4}. 
Subsequent studies of electron scattering, which were carried out in various world facilities in next decades, provide further detailed information on internal structure of various strongly interacting systems, such as nucleons, pions, and nuclei.

Because of smallness of the fine structure constant $\alpha$, theoretical interpretation of experimental data was mainly done in the lowest order in $\alpha$, or one-photon exchange (OPE) approximation (meaning the exchange of only one virtual photon between the scattered electron and the target). A fundamental ingredient of such a model is hadron electromagnetic (e.m.) current. In turn, the hadron current involves e.m. form factors (FFs), the main objects containing information on the e.m. structure of a hadron system.

Due to $1/2$ spin of the nucleon, its e.m. current is described by two FFs $F_1$ and $F_2$, called Dirac and Pauli FFs, or linear combinations of $F_1$ and $F_2$, the electric and  magnetic FFs, $G_E$ and $G_M$. To separate the FFs, two different techniques are used, the Rosenbluth separation method \cite{Rosenbluth}, which is based on the cross section data, and double polarization technique elaborated at the Thomas Jefferson National Accelerator Facility (JLab for short) \cite{Jones,Punjabi,Gayou,Puckett}. 

Up to late 1990s only the unpolarized $ep$ cross sections were measured. The Rosenbluth separation of this data had suggested that the nucleon FFs  fulfill the approximate scaling 
\be
G_{Ep} \approx G_{Mp}/\mu_p \approx G_{Mn}/\mu_n,
\ee
where $\mu_p$ and $\mu_n$ are the proton and neutron magnetic moments. Double polarization experiments carried out at JLab since 1998 has changed the situation drastically, the ratio $\mu_p G_{Ep}/G_{Mp}$ measured by this method was shown to decrease linearly with $Q^2$. Because radiative corrections for the Rosenbluth and polarization techniques are different, the discrepancy of $\mu_p G_{Ep}/G_{Mp}$ can be naturally explained as effect of radiative corrections. Theoretical analysis has shown that the two-photon exchange (TPE), contribution which was ignored in the previous calculations, may be responsible for the discrepancy \cite{bmt0,GV}.

The review is organized as follows.
Sections~\ref{sec:BA}, \ref{sec:IR}, and \ref{sec:TPEampl} have introductory purpose: we discuss methods of proton FF measurements, OPE and TPE approximations and general structure of the TPE amplitude for elastic $ep$ scattering.
Section~\ref{sec:SSA} deals with calculations of the imaginary part of TPE amplitude and applications to single spin asymmetries.
Section~\ref{sec:TPEcalc} reviews different approaches to calculation of the real part of the TPE amplitude at both hadronic and QCD levels.
Sections~\ref{sec:experiment} and \ref{sec:extraction} are devoted to discussion of the status of experimental searches for the direct TPE effects and extraction of the TPE amplitude from experimental data on the $ep$ scattering.
TPE in other processes ($\mu p$ scattering, electron scattering off the lightest nuclei, and the $e\pi$ scattering) is shortly reviewed in Sec.~\ref{sec:other}.
The Appendix contains a collection of formulae for TPE contributions to various observables.

%%%==================================================================================
\section{Born approximation and proton form factors}\label{sec:BA}

The main process of our interest will be the  elastic electron-proton scattering:
\begin{equation}
 {\rm e}^- + {\rm p} \to {\rm e}^- + {\rm p}.
\end{equation}
Though in some sections we will consider other processes, such as electron-neutron, electron-deuteron and muon-proton scattering,
by default we will mean electron-proton case, unless explicitly noted otherwise.

For the convenience, here we write down the definitions of all kinematical quantities,
which are related to this process and will be used throughout the paper.

4-momenta of the initial and final particles will be denoted $k$, $k'$
for the electron and $p$, $p'$ for the proton (the final ones are marked with the dash).
The momentum transfer is
\begin{equation}
 q = p' - p = k - k',
\end{equation}
and its square $q^2 = t = -Q^2 < 0$. Electron and proton masses will be denoted $m$ and $M$, respectively.
It is convenient to introduce vectors
\begin{equation}
 K = \tfrac{1}{2}(k+k') ,\ \ P = \tfrac{1}{2}(p+p'),
\end{equation}
for which
\begin{equation}
 Kq = Pq = 0, \ \ K^2 = m^2-t/4, \ \ P^2 = M^2-t/4.
\end{equation}
Also denote\footnote{NB: in papers of other authors $\nu$ sometimes denotes four times smaller quantity, $\nu=PK$.}
\begin{equation}
 s = (k+p)^2,\ \ u = (k'-p)^2,\ \ \nu = s-u = 4PK.
\end{equation}

\begin{figure}[h]
 \centering
 \includegraphics[width=0.225\textwidth]{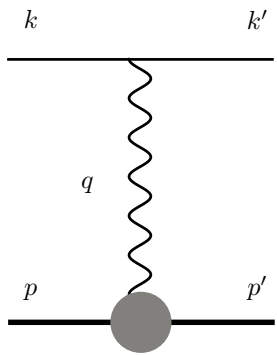}
 \caption{First-order diagram.} \label{diag1}
\end{figure}

In the first order of perturbation theory (Born or OPE approximation) electron-proton scattering is described by the only Feynman diagram (Fig.~\ref{diag1}). In this and all following diagrams thin line depicts electron, thick line --- proton, and wavy line --- photon.
Since the proton is not a point particle, the vertex, corresponding to its interaction with the virtual photon (depicted as a gray circle in Fig.~\ref{diag1}) should be written is the most general form
\begin{equation} \label{Gamma}
 \Gamma_\mu(q) = F_1(q^2) \gamma_\mu -
  \frac{1}{4M} F_2(q^2) [\gamma_\mu,\hat q],
\end{equation}
where $F_1(q^2)$ i $F_2(q^2)$ are some functions, which characterize electromagnetic interaction of the proton --- form factors (FFs). $F_1$ and is called Dirac FF and $F_2$ is called Pauli FF.
Instead of $F_1$ and $F_2$ one often uses the linear combinations
\begin{equation} \label{GEandGM}
 G_E = F_1 + \frac{q^2}{4M^2} F_2 \qquad \text{and} \qquad G_M = F_1 + F_2,
\end{equation}
--- electric and magnetic FF, respectively \cite{Sachs}.
The advantage of such a choice of FFs will become evident below.

The amplitude corresponding to the diagram Fig.~\ref{diag1}, has the form
\begin{equation} \label{ampl1}
 {\cal M}_1 = -\frac{4\pi\alpha}{q^2} \, \bar u'\gamma_\mu u \cdot \bar U' \Gamma^\mu(q) U,
\end{equation}
where $u'$, $U'$ and $u$, $U$ and bispinors of the final and initial particles, and $\Gamma_\mu$ is determined by Eq.~(\ref{Gamma}).
The cross-section of the unpolarized scattering, calculated from the amplitude (\ref{ampl1}), equals
\begin{equation} \label{Rosen}
 d\sigma = \frac{2\pi\alpha^2 dt}{E^2 t} \frac{1}{1-\ve}
  \left( \ve G_E^2 + \tau G_M^2 \right),
\end{equation}
where $\tau = -t/4M^2$, $E$ is the energy of the initial electron in the lab. frame, and
\begin{equation}\label{epsilon}
 \ve = \left[ 1 + 2 (1+\tau) \tg^2 \tfrac{\theta}{2} \right]^{-1} = \frac{\nu^2-Q^2(4M^2+Q^2)}{\nu^2+Q^2(4M^2+Q^2)},
\end{equation}
where $\theta$ is the scattering angle in that frame. The quantities $E$, $\theta$ and $\tau$ are related by
\begin{equation}
 E(E-2M\tau) = \frac{M^2\tau}{\sin^2\theta/2}
\end{equation}
The quantity $\ve$, which varies from 0 to 1,
characterizes relative contribution of the longitudinal and transverse photons to the cross-section:
the contribution of the transverse photons in independent of $\ve$,
and that of the longitudinal ones is proportional to it.

Eq.~(\ref{Rosen}), which was first derived (in somewhat different form)
in the work \cite{Rosenbluth}, is called Rosenbluth formula.
From this equation we see rationale behind the FF choice in the form (\ref{GEandGM}):
it contains only squares of the FFs and no interference terms.

Since FFs depend on $t$, but not on $\ve$, the expression
\begin{equation} \label{sigmaRed}
 \sigma_R = \ve G_E^2 + \tau G_M^2,
\end{equation}
which is called reduced cross-section, is a linear function of $\ve$ regardless of the actual FFs.

This is the basis of the {\bf Rosenbluth method} for the extraction of FFs from the experimentally measured cross-sections.
The method requires the measurements to be performed at fixed $t$ and several different values of $\ve$.
The resulting values of $\sigma_R$ are plotted against $\ve$, and should form the straight line.
Then the line slope gives us the electric FF, and the intercept gives the magnetic FF.
The linearity of the plot was, until recently, considered as a sign of the Born approximation validity.

The Rosenbluth method is very simple and was widely used for the determination of FFs of nucleons and light nuclei (such as $^3$He) from 1960s till today.
But unfortunately, it has the following drawback: the error in the electric FF rapidly increases with the momentum transfer.
Indeed, since the coefficient in front of $G_M^2$ in Eq.~(\ref{sigmaRed}) becomes large, whereas $0 \le \ve \le 1$,
the main contribution in $\sigma_R$ comes from the second term. Thus the relative error $\Delta G_E/G_E$,
which, in the order of magnitude, equals
$\Delta\sigma_R/\sigma_R \cdot \sigma_R/G_E^2 =
 \Delta\sigma/\sigma \cdot \sigma_R/G_E^2$,
will be much larger than the relative error of the cross-section $\Delta\sigma/\sigma$.

Thus, at large momentum transfers it is more convenient to use other method, {\bf the polarization transfer method}.
It was proposed in 1970s \cite{ARekalo1,ARekalo2} and first used in an experiment in 1998 \cite{Milbrath}.

This method is based on the fact that if the beam electrons are longitudinally polarized, the recoil protons become polarized as well. Their spin orientation depends on the FF ratio $G_E/G_M$, which allows to measure this ratio directly.
Namely, if the initial electrons have the helicity $\lambda$, then the polarization 4-vector of the final protons is
\begin{equation} \label{PTborn}
 S = S_\para \xi_\para + S_\perp \xi_\perp =
 \frac{-\lambda \sqrt{1-\ve^2}}{\ve G_E^2 + \tau G_M^2}
 \left( \tau G_M^2 \xi_\para + 
   \sqrt{\frac{2\ve\tau}{1+\ve}} G_M G_E \xi_\perp \right),
\end{equation}
where $\xi_\para$ i $\xi_\perp$ are unit vectors
\begin{equation}
 \xi_\para = \frac{2M}{\sqrt{-q^2 P^2}} \left( \frac{P^2}{M^2}\, p' - P \right),\ \ 
 \xi_\perp = \frac{\nu P - 4P^2 K}{\sqrt{P^2(\nu^2 + 4q^2 P^2)}},
\end{equation}
such that $\xi_\para^2 = \xi_\perp^2 = -1$, $\xi_\para p' = \xi_\perp p' = \xi_\para \xi_\perp = 0$.
Thus the FF ratio is expressed via the ratio of longitudinal and transverse polarization components:
\begin{equation} \label{Polar}
 \frac{G_E}{G_M} = \sqrt{\frac{\tau(1+\ve)}{2\ve}}
 \, \frac{S_\perp}{S_\para}
\end{equation}
Note that, contrary to Rosenbluth method, the polarization transfer method does not allow to determine $G_E$ and $G_M$ separately.
It is also more complicated technically, since it requires a polarized electron beam and a measurement of the proton polarization.
On the other hand, since only the {\it ratio} of the polarization components is measured, where is no need to know beam polarization of analyzing power of the detector exactly.
The main advantage of the method is that the ratio $G_E/G_M$ is determined with higher accuracy, especially at large momentum transfers.

From the theoretical point of view, the equivalent to polarization transfer method is {\bf beam-target asymmetry method} \cite{PTJones},
though the experimental setup is quite different. Longitudinally polarized electrons are scattered off the polarized protons, and one observes the asymmetry
$A = \frac{\sigma_+ - \sigma_-}{\sigma_+ + \sigma_-}$,
where $\sigma_{\pm}$ is the cross-section for the scattering of electrons with the helicity $\pm\half$.

In Born approximation, $A$ is also expressed via the FF ratio.
Using time reversal symmetry one can show that the results obtained with this method should be identical to the results of polarization transfer method, even beyond the Born approximation.
Thus further, for brevity, we will speak of the polarization transfer method, implying the beam-target asymmetry method as its special case.

\begin{figure}[h]
 \centering \includegraphics[width=0.75\textwidth]{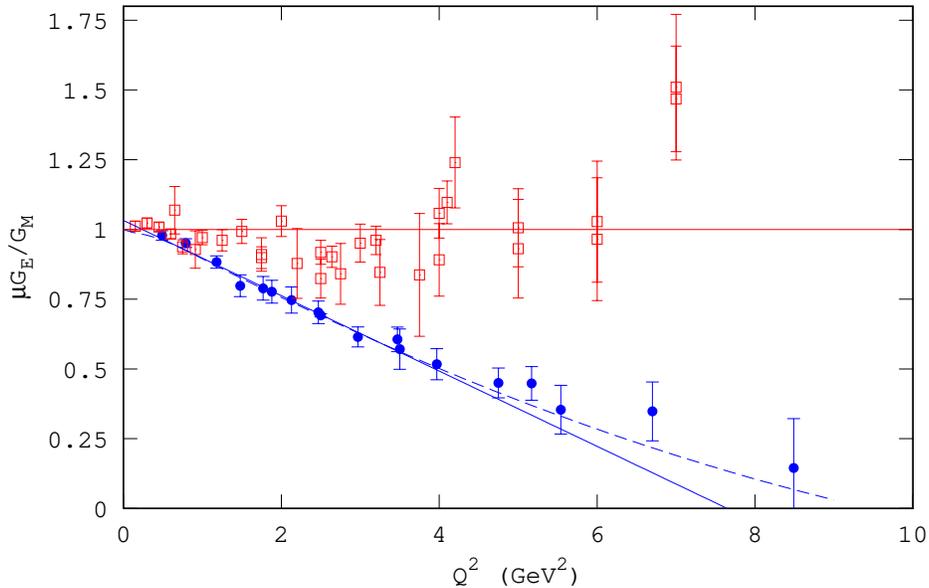}
 \caption{Experimentally measured proton FF ratio. Squares --- Rosenbluth method \cite{LTWalker,LTAndivahis,LTChristy,LTQattan},
circles --- polarization transfer method \cite{Punjabi,Puckett,Puckett2017,GEp2gamma}. Blue line --- Eq.(\ref{linear}), dashed --- parameterization from \cite{ArringtonFit}.}
 \label{ge/gm}
\end{figure}

Proton FFs, obtained by the Rosenbluth method, approximately obey the relation $\mu G_E/G_M \approx 1$, where $\mu $ is the proton magnetic moment.
First polarization transfer measurements were performed at small $Q^2$ and confirmed this relation \cite{Milbrath}.
But when the momentum transfer values were increased up to $Q^2 \gtrsim 1 \GeV^2$,
results become unexpected: the ratio $G_E/G_M$ decreased monotonically as $Q^2$ increased and
obviously disagreed with values obtained by Rosenbluth method.
Since then, several other experiments were performed by both methods, in which the accuracy was improved and higher momentum transfers reached ($Q^2$ up to $8.5\GeV^2$, Fig.~\ref{ge/gm}).
The values of $G_E/G_M$, obtained by different authors by the same method, agree with each other well, whereas the values obtained by different methods, significantly disagree. In particular, the ratio $G_E/G_M$, measured by the polarization transfer method, at not very high $Q^2$ was well-described by a linear function \cite{Arrington}
\begin{equation} \label{linear}
 \mu G_E/G_M = 1 - 0.135 (Q^2 - 0.24)
\end{equation}
(solid line in Fig.~\ref{ge/gm}).

One should note that such an asymptotic behaviour of the FFs contradict the results obtained in the framework of QCD. Perturbative QCD calculations yield the dependence $F_1 \sim Q^2 F_2$ at large $Q^2$, which is equivalent to $G_E/G_M \sim \rm const$ \cite{QCD1,QCD2}.

More accurate statistical analysis confirmed the above-mentioned statements.
In particular, it was shown in Ref.~\cite{Arrington} that
\begin{itemize}
\item the cross-sections, measured in Rosenbluth method, do not contain rough errors
\item the cross-section, measured in different experiments, are consistent with each other
\item there is no statistically sound parameterization of FFs, compatible with the results obtained by both methods
\end{itemize}
This led to the suggestion that, if we exclude a possibility of the rough error in the polarization transfer measurements, the discrepancy is likely caused by the terms of the next (second) order of the perturbation theory, which were neglected when deriving Eqs. (\ref{sigmaRed}) and (\ref{Polar}).

%%%==================================================================================
\section{Second order perturbation theory} \label{sec:IR}
\begin{figure}[b]
\centering
$\begin{array}{cccc}
 \includegraphics[width=0.16\textwidth]{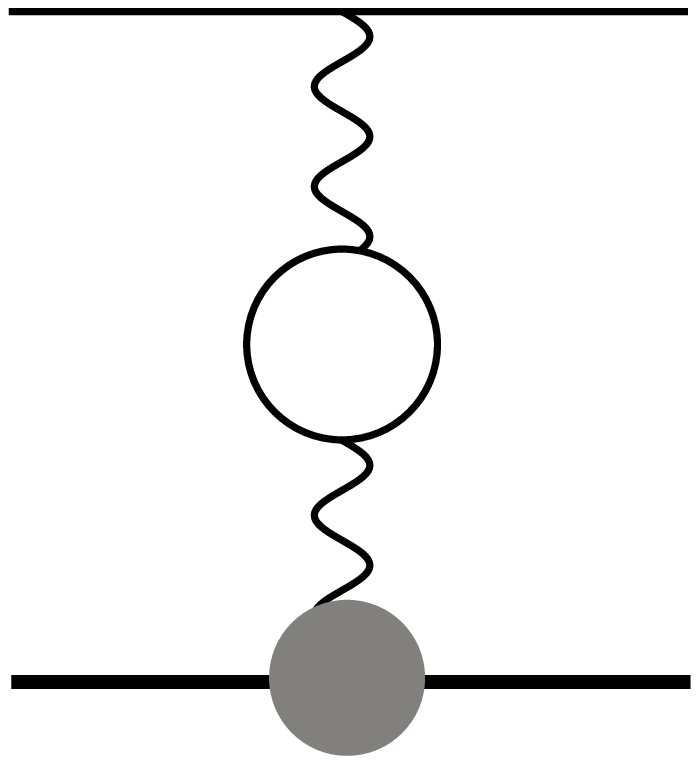} &
 \includegraphics[width=0.16\textwidth]{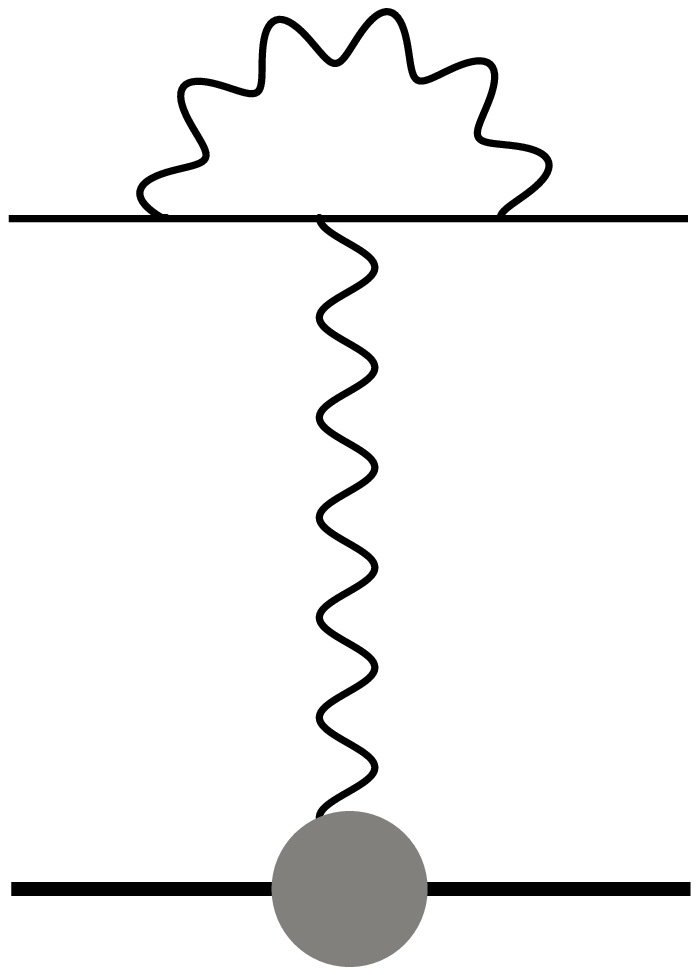} &
 \raisebox{-2.5mm}{\includegraphics[width=0.2\textwidth]{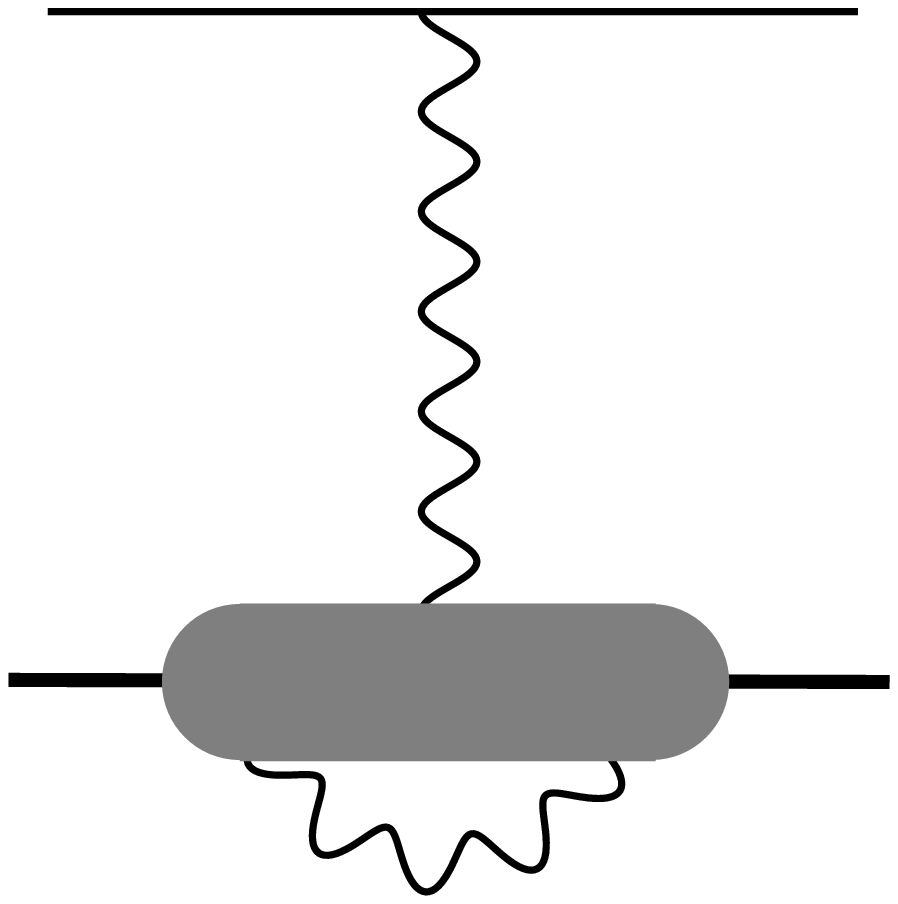}} &
 \includegraphics[width=0.2\textwidth]{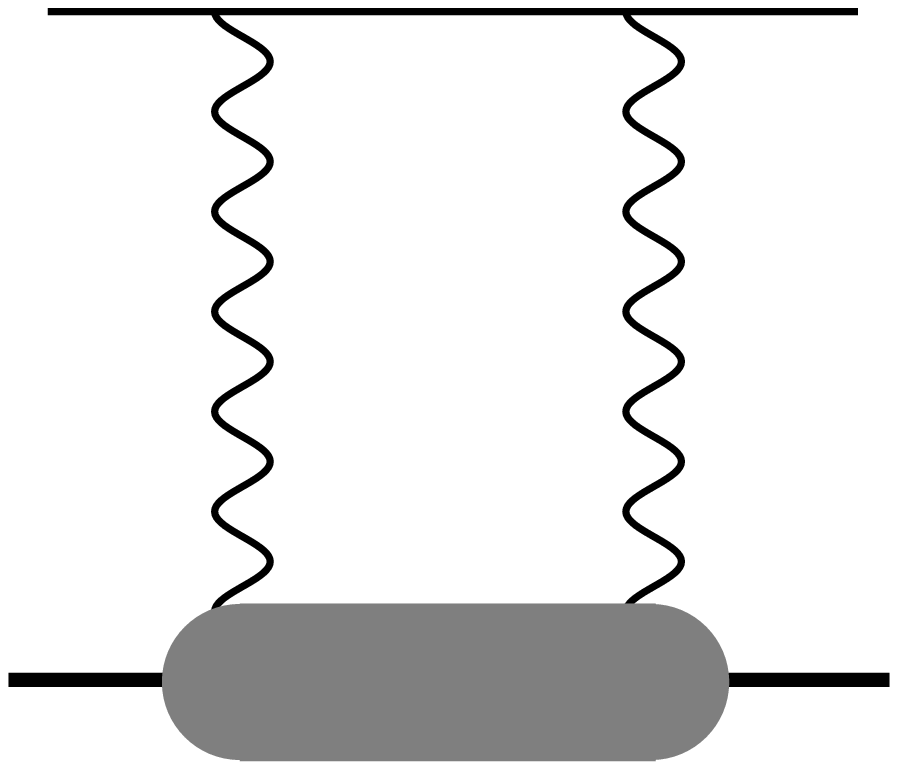} \\
 $(a)$ & $(b)$ & $(c)$ & $(d)$
 \end{array}$
 \caption{Second order diagrams.} \label{diag2}
\end{figure}

In the second order of the perturbation theory, several Feynman diagrams exist (Fig.~\ref{diag2}):
vacuum polarization (\ref{diag2}a), electron-photon and proton-photon vertex corrections
(\ref{diag2}b and \ref{diag2}c, respectively) and TPE diagram (\ref{diag2}d).
We do not draw a diagram, analogous to \ref{diag2}d, with crossed photon lines, since we treat its lower part as already symmetrized with respect to photon interchange.

These diagrams have some new properties, which were absent in the Born approximation.

First, there is so-called infra-red (IR) divergence.
The integrals, corresponding to diagrams Fig.~\ref{diag2}b,c,d, are logarithmically divergent when virtual photon momenta go to zero.
Thus, taking into account these diagrams, the elastic scattering cross-section becomes infinite.
It is well-known that this stems from the incorrect formulation of the problem.
Since any detector has finite energy resolution, it is impossible for the experiment to separate
the elastic process from the process with additional emission of any number of soft photons with a total energy less than some threshold.

Thus the true observable is only $\sigma(\Delta E)$ --- the cross-section of such a process in which the sum of final electron and proton energies differs from the initial energy by not more than some $\Delta E > 0$.

This cross-section is convenient to calculate in the following way.
Suppose that the photon has small but non-zero mass $\la$.
Then all IR-divergent integrals become finite, but instead contain terms, proportional to $\ln \la$.
The cross-section $\sigma(\Delta E)$ is a sum of two quantities: the elastic scattering cross-section $\sigma_{el}$
and the soft photon emission cross-section $\sigma_\gamma(\Delta E)$.
Either of these quantities is divergent at $\la\to 0$, but in sum the divergent terms exactly cancel,
and the quantity $\sigma(\Delta E)$ appears well-defined at $\la\to 0$.

The form of the IR-divergent terms can be determined without calculation of the scattering amplitude and even without the expansion in $\alpha$ \cite{Landau}. 
Indeed, if the quantity $\Delta E$ is small, then the photon emission follows classical electrodynamics,
and the photons are emitted independently of each other.
The number of these photons will be Poisson-distributed random quantity; the probability of the scattering with emitting exactly
$n$ photons equals
\[
 w_n = \frac{1}{n!}w^n \exp(-w),
\]
where $w$ is the probability of a photon emission:
\begin{equation}
 w = \frac{\alpha}{\pi} \left( B_1 \ln \frac{\Delta E}{\la} + B_2 \right),
\end{equation}
where $B_1$ and $B_2$ --- some functions of the particle momenta, which are known explicitly.
So, we have a relation between the cross-section of the scattering with the emission of arbitrary number of photons $\sigma(\Delta E)$
and the cross-section of the ``elastic'' scattering, i.e. without any emission $\sigma_{el}$:
\begin{equation}
 \sigma_{el} = w_0 \sigma(\Delta E) = \exp(-w) \sigma(\Delta E),
\end{equation}
or
\begin{equation}
 \sigma(\Delta E) = \sigma_{el} \exp \left[ \frac{\alpha}{\pi}
   \left( B_1 \ln \frac{\Delta E}{\la} + B_2 \right) \right]
\end{equation}
We see that for the IR-divergencies to cancel, the quantity $\sigma_{el}$ must have the form
\begin{equation} \label{IRsigma}
 \sigma_{el} = \sigma_0 \exp \left[ \frac{\alpha}{\pi}
   \left( B_1 \ln \frac{\la}{E} + B_3 \right) \right]
\end{equation}
where $\sigma_0$ is independent of $\la$.
Then
\begin{equation}
 \sigma(\Delta E) = \sigma_0 \exp \left[ \frac{\alpha}{\pi}
   \left( B_1 \ln \frac{\Delta E}{E} + B_2 + B_3 \right) \right]
\end{equation}
The auxilliary quantity $\la$ disappears from the formulae, but instead the cross-section becomes dependent on $\Delta E$.
The cross-section of exactly elastic scattering, i.e. $\lim\limits_{\Delta E\to 0} \sigma(\Delta E)$, is zero.
This is clear from the physical grounds: every collision of charged particles is accompanied by the emission of the electromagnetic waves, that is, soft photons.

Expanding (\ref{IRsigma}) in series in $\alpha$, one can find the IR-divergent terms in $\sigma_{el}$
in every order of the perturbation theory.
They cancel with similar terms in the soft photon emission cross-section $\sigma_\gamma(\Delta E) = \sigma(\Delta E)-\sigma_{el}$.

\begin{figure}[h]
 \centering
 \includegraphics[width=0.4\textwidth]{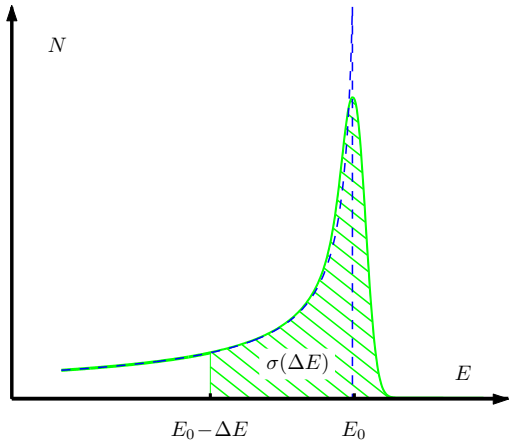} 
 \caption{Number of the events $N$ depending on the final particles energy $E$
in the elastic scattering experiment.} \label{figsig}
\end{figure}

Then, what quantity is really measured in the experiments on elastic scattering?
Aside from minor details, which can differ between experiments, the elastic cross-section measurement is performed in the following way.
For every scattering event the energy $E$ of the final particles (the electron and the proton) is determined.
The typical histogram of $E$ is shown in Fig.~\ref{figsig},
$E_0$ is the initial energy. ($E$ can be greater then $E_0$ because of the beam non-monochromaticity, finite detector resolution and so on. The ideal experiment would yield the dashed curve; the solid curve,
obtained in the real experiment is a convolution with a sort of ``instrument function'').
Then one chooses some $\Delta E$ and all events with $E>E_0-\Delta E$ are formally counted as elastic.
The corresponding cross-section $\sigma(\Delta E)$ is proportional to the hatched area in Fig.~\ref{figsig}
and is called ``uncorrected'' or ``measured'' cross-section.
Then the experimenters, starting from the measured quantity $\sigma(\Delta E)$,
try to calculate the elastic scattering cross-section {\it in Born approximation},
which is finally published as ``the elastic cross-section with radiative corrections'' (``corrected cross-section'') $\sigma_{cor}$.
The quantity $\sigma(\Delta E) - \sigma_{cor}$ is called radiative correction and includes both the cross-section of the soft photon emission and the higher-order corrections to the elastic cross-section $\sigma_{el}$.
It can reach 20-30\% (see, e.g. \cite{LTWalker}).

It is clear that the radiative corrections are to be calculated theoretically.
The standard procedure for this calculation was published in Ref.~\cite{MoTsai}
and is based on the results of Ref.~\cite{Tsai}.
Most of the experimental works, especially older ones, used this procedure.
Let us consider it in some detail.

For the vacuum polarization and electron-photon vertex correction (\ref{diag2}a i \ref{diag2}b)
one uses the exact expressions, obtained in QED \cite{Landau}:
\begin{equation}
 \CM_{\text{2a}} = f_{\text{a}}(q^2) \CM_1,\ \
 \CM_{\text{2b}} = f_{\text{b}}(q^2) \CM_1,
\end{equation}
where
\begin{equation} \label{M2a}
 f_{\text{a}}(q^2) = \frac\alpha{3\pi} \ln \frac{-q^2}{m^2}
\end{equation}
for the diagram \ref{diag2}a and
\begin{equation} \label{M2b}
 f_{\text{b}}(q^2) = -\frac\alpha{2\pi} \left( \frac12 \ln^2 \frac{-q^2}{m^2} +
  2 \ln\frac{m}\la \ln \frac{-q^2}{m^2} \right)
\end{equation}
for the diagram \ref{diag2}b (supposing that $-q^2 \gg m^2$).
Due to presence of the large logarithm $\ln\frac{-q^2}{m^2}$ in the expressions (\ref{M2a}, \ref{M2b}),
these corrections may be quite large; in some high-energy experiments \cite{LTWalker}
muon and hadron vacuum polarization and higher order corrections to the electron-photon vertex were also taken into account.

The diagrams \ref{diag2}c and \ref{diag2}d in Refs.~\cite{MoTsai,Tsai} were calculated in so-called soft photon approximation (or Mo-Tsai approximation), that is, supposing that the momentum of one of the virtual photons is close to zero.
This way one can exactly determine IR-divergent contribution, which must cancel the same term in the inelastic cross-section.

However, even if we forget the above, the diagrams \ref{diag2}a-\ref{diag2}c
cannot cause the discrepancy between Rosenbluth and polarization transfer methods in the measurements of the FFs.
The point is that they all has the structure analogous to the Born amplitude, namely
\begin{equation}
 \CM = -\frac{4\pi\alpha}{q^2} \bar u'\gamma_\mu u \cdot \bar U' \delta\Gamma^\mu(q) U,
\end{equation}
where
\begin{equation}
 \delta\Gamma_\mu(q) = \delta F_1(q^2) \gamma_\mu -
  \frac{1}{4M} \delta F_2(q^2) [\gamma_\mu,\hat q].
\end{equation}
Thus inclusion of these diagrams does not break the formulae obtained in the Born approximation,
but leads only to the effective change $F_i \to F_i + \delta F_i$.
Thus, though the results, obtained by both methods, may change, the change will be the same and the discrepancy cannot arise.

So we end up with the conclusion, the the only non-trivial diagram, which can be responsible for the discrepancy, is TPE diagram (Fig.~\ref{diag2}d).
The effect of TPE in three-way:
\begin{itemize}
\item first, now the amplitude has non-zero imaginary part,
which gives rise to such effects as single-spin asymmetries (absent in OPE)
\item second, there is a correction to the real part of the amplitude,
which now has different tensor structure than in OPE, consequently breaking Rosenbluth formula.
\item the TPE correction has opposite signs for $e^-p$ and $e^+p$ scattering, leading to the {\it charge asymmetry}: the cross-section for electron and positron scattering are now different.
\end{itemize}

Note however, that in soft-photon approximation of Mo-Tsai the diagram Fig.~\ref{diag2}d has the same factorization property as other three: it is proportional to the OPE amplitude:
\be
 \CM_\text{2d}^\text{(Mo-Tsai)} = f_\text{d}(\nu, q^2) \CM_1
\ee
Thus, non-trivial TPE effects come not just from TPE diagram, but from its IR-finite, hard-photon part $\CM_\text{2d} - \CM_\text{2d}^\text{(Mo-Tsai)}$.

\begin{figure}[h]
\centering
$\begin{array}{cccc}
 \includegraphics[width=0.15\textwidth]{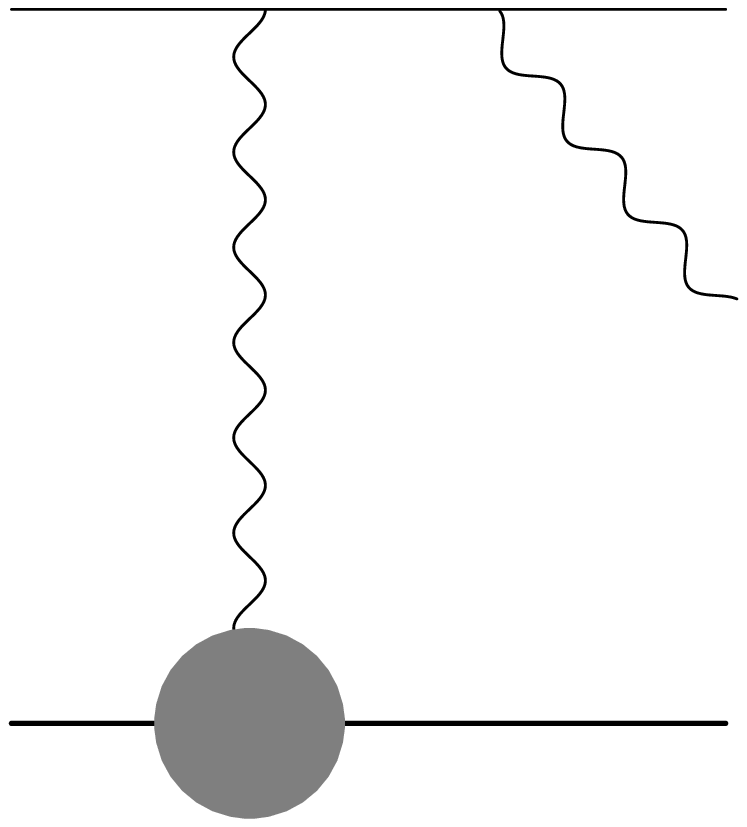}\quad &
 \includegraphics[width=0.15\textwidth]{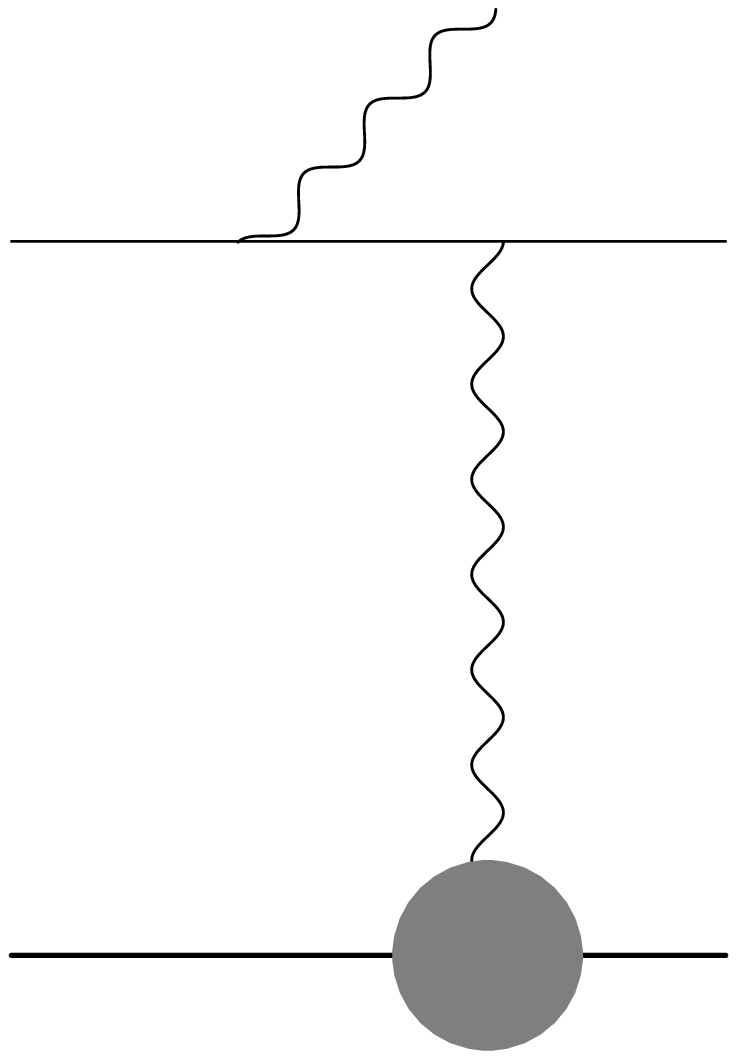}\quad &
 \includegraphics[width=0.15\textwidth]{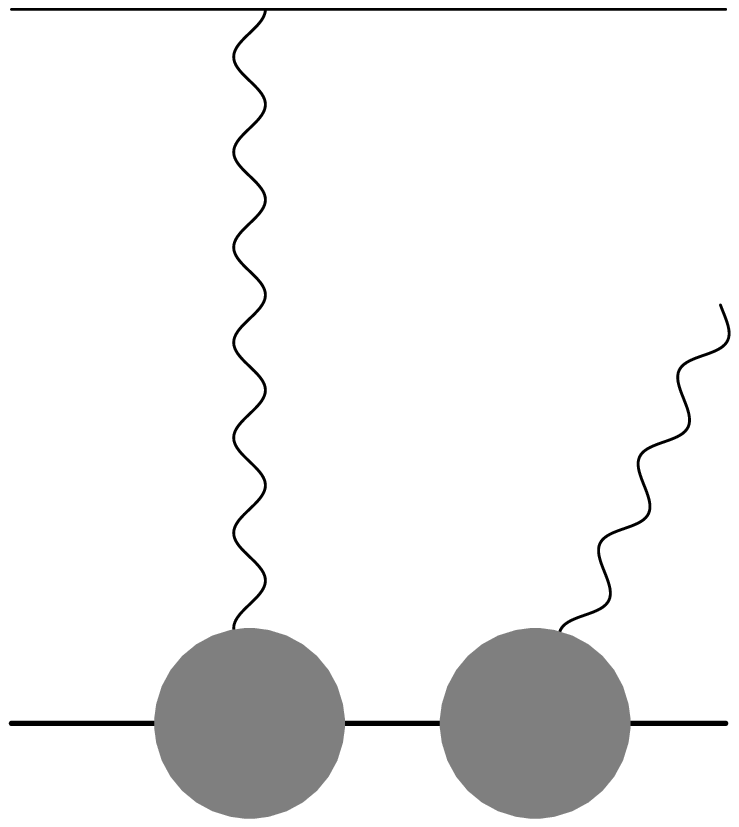}\quad &
 \includegraphics[width=0.15\textwidth]{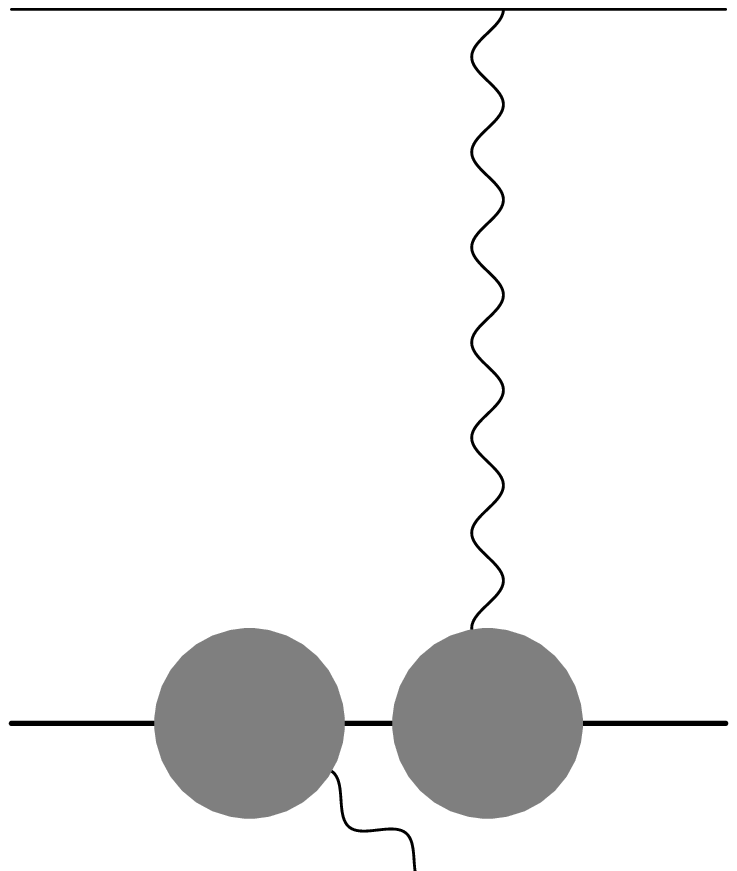} \\
 $(a)$ & $(b)$ & $(c)$ & $(d)$
 \end{array}$
 \caption{Photon emission (bremsstrahlung) diagrams.} \label{diag-bs}
\end{figure}

Another component that could contribute to the discrepancy between Rosenbluth and polarization transfer methods,
is higher order (in $\Delta E$) corrections to bremsstrahlung cross-section $\sigma_\gamma(\Delta E)$,
which, in the leading order in $\alpha$, is described by the diagrams Fig.~\ref{diag-bs}.
This was addressed in Refs.\cite{AfaBS1,AfaBS2}, where the radiation by the electron (Fig.~\ref{diag-bs}a,b)
was thoroughly studied in a model-independent way;
and in Ref.\cite{ourBS}, where the radiation by the proton was studied in next-to-leading order in $\Delta E$,
and it was shown that, at least at high $Q^2$, these corrections are smaller than TPE and do not influence the experimental results noticeably.
Thus further we will concentrate on TPE corrections only.
%
%%%==================================================================================
\section{Structure of the TPE amplitude}\label{sec:TPEampl}
In general case, the elastic scattering of the two non-identical particles with spin 1/2,
such as electron and proton,
is described by six scalar functions --- invariant amplitudes,
which may be taken as, for example, helicity amplitudes \cite{Landau}.
However, the typical energies involved in the electron-proton scattering experiments
are several order of magnitude greater than the electron mass, thus the latter can be safely neglected.
Then the electron helicity is conserved, and the number of amplitudes decreases to three, which can be chosen as \cite{GV}:
\begin{equation} \label{genAmp}
 {\cal M} = -\frac{4\pi\alpha}{q^2} \bar u'\gamma_\mu u \cdot
 \bar U' \left(\gamma^\mu \tilde F_1
 - \frac{1}{4M} [\gamma^\mu,\hat q] \tilde F_2
 + \frac{P^\mu}{M^2} \hat K \tilde F_3
 \right) U.
\end{equation}
The invariant amplitudes, or generalized FFs, $\tilde F_i$ are functions of the two kinematical variables,
$\nu = 4PK$ and $t=q^2$. In the Born (OPE) approximation, the dependence on $\nu$ disappears, and the amplitudes reduce to the usual FFs:
\[
\begin{array}{l}
\tilde F_1^\text{(Born)}(\nu,t) = F_1(t),\\
\tilde F_2^\text{(Born)}(\nu,t) = F_2(t),\\
\tilde F_3^\text{(Born)}(\nu,t) = 0
\end{array}
\]
(compare (\ref{genAmp}) with (\ref{Gamma}) and (\ref{ampl1})).
The TPE diagram gives a contribution of order $O(\alpha)$ to each of these amplitudes.
It is clear that one can choose any three independent linear combinations of $\tilde F_i$ as invariant amplitudes.
Some authors introduce, in analogy with the electric and magnetic FFs, the quantities
\[
\tilde G_M = \tilde F_1+\tilde F_2 \ \text{ and }\ \tilde G_E = \tilde F_1 - \tau \tilde F_2.
\]
However the most convenient choice of the amplitudes is, to our opinion, the following:
\begin{equation} \label{amplCG}
 \begin{array}{rclll}
  \CG_E & = & \tilde F_1 - \tau \tilde F_2 + \nu \tilde F_3/4M^2 &\qquad & \CG_E^\text{(Born)} = G_E\\
  \CG_M & = & \tilde F_1 +\tilde F_2 + \ve \nu \tilde F_3/4M^2 & & \CG_M^\text{(Born)} = G_M\\
  \CG_3 & = & \nu \tilde F_3/4M^2 & & \CG_3^\text{(Born)} = 0
 \end{array}
\end{equation}
Then, if the scattering amplitude has the general form (\ref{genAmp}),
the unpolarized cross-section becomes
\begin{equation} \label{Rosen2g}
 d\sigma = \frac{2\pi\alpha^2 dt}{E^2 t} \frac{1}{1-\ve}
 \left( \ve |\CG_E|^2 + \tau |\CG_M|^2 + \tau \ve^2 \frac{1-\ve}{1+\ve} |\CG_3|^2 \right)
\end{equation}
This formula is exact and do not rely on the expansion in $\alpha$.
We see that, in analogy with the Rosenbluth formula, there is no interference terms.
Eq.~(\ref{PTborn}) for the final proton polarization in the polarization transfer method turns to
\begin{equation} \label{PT2g}
 S = \frac{-\lambda \sqrt{1-\ve^2}}{\ve |\CG_E|^2 + \tau |\CG_M|^2}
 \Re \tilde G_M^* 
 \left[ \left( \CG_M+\tfrac{\ve(1-\ve)}{1+\ve}\CG_3 \right) \tau \xi_\para + 
   \sqrt{\tfrac{2\ve\tau}{1+\ve}} \CG_E \xi_\perp \right]
\end{equation}
Contrary to the Born approximation, in general case the amplitudes $\CG$ are complex-valued.
Their imaginary part is proportional to $\alpha$ and comes exclusively from the TPE diagram.
On the other hand, the equations (\ref{Rosen2g}, \ref{PT2g}), neglecting terms of order $O(\alpha^2)$,
contain only their real parts.
This follows from the fact that the Born amplitudes are real; e.g.,
\[ |\CG_E|^2 = (G_E + \Re\delta\CG_E)^2 + (\Im\delta\CG_E)^2 =
 G_E^2 + 2G_E\Re\delta\CG_E + O(\alpha^2), \]
where $\delta\CG_E$ is the TPE amplitude.
The imaginary part of the amplitudes gives rise to the new type of observables --- 
single-spin asymmetries.

In the first order of the perturbation theory, for the unpolarized scattering
the final particles are unpolarized as well.
If only one of the initial particles is polarized, the scattering cross-section is independent of its spin
and still follows Eq.~(\ref{Rosen}).
With taking into account TPE, such dependence arises.
The situation where the spin of polarized particle is perpendicular to the reaction plane, is of special interest.
Since in this case it has only two possible directions (say, up and down)
then, denoting corresponding cross-sections as $\sigma_\uparrow$ and $\sigma_\downarrow$,
we define the asymmetry as
\begin{equation} \label{defAsym}
 A_n,B_n = \frac{\sigma_\up - \sigma_\down}{\sigma_\up + \sigma_\down}.
\end{equation}
When the polarized particle is proton (the target) this quantity is called ``target normal spin asymmetry" (TNSA),
and denoted $A_n$, and when it is electron it will be denoted $B_n$ and called
``beam normal spin asymmetry" (BNSA).

The interesting properties of these quantities are:
\begin{itemize}
\item as it was stated above, both asymmetries are strictly zero in the OPE approximation, thus the leading contribution to the asymmetry is given by TPE.
\item they are expressed via the imaginary part of the scattering amplitude, which significantly simplifies theoretical calculations.
\item the asymmetry does not contain IR divergencies. This can be inferred immediately from the definition (\ref{defAsym}):
 both cross-sections, $\sigma_\uparrow$ and $\sigma_\downarrow$,
 contain the same IR-divergent factor (the exponent from Eq.~(\ref{IRsigma})),
 which thus cancels. Therefore, there is no need for radiative corrections in the measurements of the asymmetry.
\end{itemize}

TNSA can be expressed via the amplitudes $\CG$, introduced above, as
\begin{equation}
 A_n = \frac{\sqrt{2\tau\ve(1+\ve)}}{\tau G_M^2+\ve G_E^2}
 \left[ G_E \Im \left( \CG_M+\tfrac{\ve(1-\ve)}{1+\ve}\CG_3 \right) -
    G_M \Im \CG_E \right] + O(\alpha^2).
\end{equation}
BNSA cannot be expressed via these amplitudes, since it vanishes for the zero electron mass.
Thus it also depends on the other three amplitudes, which were neglected in Eq.~(\ref{genAmp}).

Single-spin asymmetries are considered in detail in Sec.~\ref{sec:SSA}.

Calculation of the whole TPE amplitude (and not just its imaginary part) is a difficult task.
Early works, considering TPE for the electron-proton scattering \cite{DrellFu,Greenhut}, has the following common drawbacks:
\begin{itemize}
 \item usually the proton is considered as a source of constant external field
 (static approximation, $M\to\infty$).
 \item only cross-section correction is calculated but not the scattering amplitude, which has three independent components (\ref{genAmp})
\end{itemize}
The main difficulties with the rigorous calculation of the TPE amplitudes are the need to
take into account large number of the intermediate states and the FFs in the $\gamma^*p$ vertices.

%%%==================================================================================
\section{Imaginary part effects} \label{sec:SSA}

\begin{figure}[b] 
 \centering
 \includegraphics[width=0.5\textwidth]{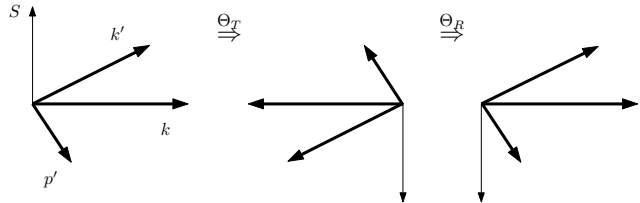}
 \caption{To the derivation of formula for TNSA}\label{transform}
\end{figure}

Let us demonstrate that the single-spin asymmetry (say, TNSA) is proportional to the imaginary part of the TPE amplitude \cite{deRuj1}.
Recall that TNSA is defined as
\begin{equation}
A_n = \frac{\sigma_\uparrow - \sigma_\downarrow}
         {\sigma_\uparrow + \sigma_\downarrow}
\end{equation}
where $\sigma_\uparrow$ and $\sigma_\downarrow$ are cross-sections for the unpolarized electron scattering
off the polarized proton with the spin perpendicular to the scattering plane,
directed either up ($\sigma_\uparrow$) or down ($\sigma_\downarrow$).
Denote the initial state $|i\>$, and the final state $|f\>$;
the same states with the spins reversed are marked by a hat: $|\hat i\>$ and $|\hat f\>$.
Then the cross-sections of our interest are, up to a constant factor
\begin{equation}
  \sigma_\up = \sum |T_{fi}|^2, \ \ \ \ 
  \sigma_\down =
    \sum |T_{f \hat i}|^2 = \sum |T_{\hat f \hat i}|^2,
\end{equation}
where $\sum$ mean summation over the spins of the initial and final electron and the final proton.
Acting on the states $|i\>$ and $|f\>$
by time reversal ($\Theta_T$) and rotation around the normal to the scattering plane by $180^\circ$ ($\Theta_R$), see Fig.~\ref{transform}, we have
\begin{equation}
  \Theta_R \Theta_T |i\> = \eta_i |\hat i\>,\ \ \
  \Theta_R \Theta_T |f\> = \eta_f |\hat f\>,
\end{equation}
where $\eta_{i,f}$ are some phases, $|\eta_{i,f}|^2=1$.
From $T$-invariance of the electromagnetic interaction it follows that
\begin{equation} \label{T-inv}
  T_{\hat f\hat i} = \eta_i \eta_f^* T_{if}.
\end{equation}
With the help of Eq.(\ref{T-inv}) the cross-section difference can be written as
\begin{eqnarray} \label{sig:up-dn}
  & & \sigma_\up - \sigma_\down = \sum \left[ |T_{fi}|^2 - |T_{if}|^2 \right] = \\
  & & = \frac{1}{2} \sum \left[ (T_{fi}^* + T_{if})(T_{fi}-T_{if}^*) -
        (T_{if}^* + T_{fi})(T_{if}-T_{fi}^*) \right] = \nonumber \\
  & & = \frac{1}{2} \sum \left[ (T_{fi}^* + T_{if})(T_{fi}-T_{if}^*) -
        (T_{f\hat i}^* + T_{\hat i f})(T_{f\hat i}-T_{\hat i f}^*) \right].
      \nonumber 
\end{eqnarray}
The unitarity condition says
\begin{equation}
 T_{fi} - T_{if}^* = i \sum_n T_{fn} T_{in}^*,
\end{equation}
where the summation goes over the complete set of intermediate states $n$.
In the case of electromagnetic interaction $T$-matrix elements are proportional to the small quantity $\alpha$.
This means that in the first order in $\alpha$ $T$-matrix is hermitian:
\begin{equation}
 T_{fi}^{(1)} = T_{if}^{(1)*},
\end{equation}
and the antihermitian part (which is usually somewhat inaccurately called ``imaginary'')
arises only in the second order:
\begin{equation} \label{unit1}
 T_{fi} - T_{if}^* = i \sum_n T_{fn}^{(1)} T_{in}^{(1)} + O(\alpha^3).
\end{equation}
Inserting (\ref{unit1}) in (\ref{sig:up-dn}), we see that in the first non-vanishing order in $\alpha$ the asymmetry has the form
\begin{eqnarray} \label{asy1}
 A_n \approx i \sum_n \frac{\sum T_{fn}^{(1)}\left(
   T_{ni}^{(1)} T_{if}^{(1)} - T_{n\hat i}^{(1)} T_{\hat if}^{(1)}
    \right)}{\sum \left( |T_{fi}^{(1)}|^2 + |T_{f\hat i}^{(1)}|^2 \right)}
\end{eqnarray}
Of course, similar equation holds for the BNSA (when the polarized particle is electron).

The outer summation in Eq.(\ref{asy1}) goes over all states $n$ that can be produced in the interaction of the electron with the proton.
In the first order in $\alpha$ they must consist of the electron and some hadronic state with baryon number 1, charge +1, and invariant mass $W \le \sqrt{s}$.
This hadronic state can be bare proton, proton plus some number of mesons ($\pi p$, $\pi\pi p$, $\eta p$...) or proton-antiproton pairs ($pp\tilde p$) etc.
The contribution of the bare proton as the intermediate hadronic state is called the elastic contribution
(as the production of this state would be the elastic process); other contributions are called inelastic.

The main problem in calculations of the asymmetry comes from the inelastic contribution, since at not-so-small energies
the number of allowed intermediate states can be large, therefore one has to use some model for this states.

\subsection{TNSA}
The TNSA was first considered in Refs.~\cite{deRuj1, deRuj2}.
The elastic part was calculated directly and the inelastic part was constrained using a sort of Cauchy inequality:
\be \label{deRuj-bound}
\left| T_{fi} - T_{if}^* \right| = 
\left| \sum_n T_{fn} T_{in}^* \right| \le
\sqrt{ \sum_n \left| T_{fn} \right|^2 \sum_n \left|T_{in}\right|^2} \sim \sqrt{\sigma_{f} \sigma_{i}}
\ee
where $\sigma_i$ ($\sigma_f$) is total inelastic cross-section from the state $i$ ($f$).

The advantage of such an approach is that the above expression includes only experimentally measurable quantities,
thus the constraint is model-independent. However, as it was already noted by the authors,
while for the near-forward scattering the bound given by the Eq.(\ref{deRuj-bound}) is likely a good approximation to the imaginary part of the amplitude,
for the large-angle scattering it probably far overestimates its true value.
Thus we need to do more accurate estimate of the inelastic contribution to the TNSA.

Such calculations were performed in Refs.~\cite{PV,ourTNSA}.
In Ref.~\cite{PV} the intermediate states included in the unitarity relation were $\pi N$ states.
These are lowest inelastic states possible (other inelastic states include $\pi\pi N$, $\pi\pi\pi N$, $\eta N$ and so on, and their contribution was neglected).
The amplitudes of the pion electroproduction $\gamma^* N \to \pi N$, which enter unitarity condition in this case,
are reasonably well-known and were taken from the MAID model \cite{MAID}.

In Ref.~\cite{ourTNSA} the lowest resonances were used instead, — namely, $P_{33}(1232)$, $D_{13}(1520)$,
$S_{11}(1535)$, $F_{15}(1680)$ and $P_{11}(1440)$, with transition amplitudes fitted from the experimental data.
Compared to Ref.~\cite{PV}, on one hand, we include some states other than $\pi N$,
since the resonances are not 100\% composed of $\pi N$, but on the the other hand we leave out nonresonant pion production.

It was found that the contributions of the resonances tend to cancel each other, and that for the larger energies the asymmetry
(and the imaginary part of the TPE amplitude as well) is dominated by the elastic part.
Since the real and imaginary parts of the amplitude are connected via dispersion relations,
it was suggested that the elastic contribution is a good approximation for the real part of the TPE amplitude as well.

\subsection{BNSA}

The BNSA is proportional to the lepton mass and vanishes in the $m\to 0$ limit.
Indeed, a state with the electron spin perpendicular to the scattering plane is the superposition of the states with positive and negative helicity,
with the coefficients differing only in phase. But the helicity of a massless particle is conserved, thus there is no interference between different helicities and the cross-sections $\sigma_\up$ and $\sigma_\down$ are equal.

However, numerically the asymmetry is not as small as one may expect, since it contains logarithmic ($\ln Q^2/m^2$) and double-logarithmic ($\ln^2 Q^2/m^2$) enhancements,
as it was shown in Refs.~\cite{AfanMerCorr, Gorchtein, ourBNSA}.
The interplay between these terms is such that $\ln Q^2/m^2$ term dominates at forward angles and the $\ln^2 Q^2/m^2$ term at backward ones,
because of different powers of $Q^2$ in front of the logarithms.

The following representation of BNSA can be easily derived from Eq.(\ref{asy1}):
\begin{equation}
  B_n = \frac{i\alpha q^2 m}{2\pi^2 D}
   \int \frac{d^3 k''}{2\epsilon''} \frac{1}{q_1^2 q_2^2} Y(W,q_1^2,q_2^2) + o(m),
\end{equation}
where function $Y(W,q_1^2,q_2^2)$ is a contraction of leptonic and hadronic tensors describing virtual Compton scattering, whose its exact form is not needed here, $W = (p+k-k'')^2$ is the invariant mass of the hadronic intermediate state, and
$D = \frac{ 4(2s + q^2 - 2M^2)^2}{4 M^2 - q^2} (4 M^2 G_E^2 - q^2 G_M^2) + 4 q^2 (4 M^2 G_E^2 + q^2 G_M^2)$.
%%
%\begin{equation}
% Y(W,q_1^2,q_2^2) = \tilde L^{\alpha\mu\nu}
% \sum_{\la_p,\la'_p}
% W_{\mu\nu}(p'\la'_p;p\la_p)\ \<p\la_p| \Gamma_\alpha |p'\la'_p\>.
%\end{equation}
%%
The double-logarithmic terms, found in Ref.~\cite{ourBNSA}, arise form the approximate formula
\be \label{int}
 \int \frac{d^3 k''}{2\epsilon''}\frac{1}{q_1^2 q_2^2} Y(W,q_1^2,q_2^2) \approx 
 \frac{\pi}{4 Q^2} Y(\sqrt{s}, 0, 0) \ln^2 \frac{Q^2}{m^2},
\ee
which yields
\be \label{Bn}
 B_n \approx B_n^{(\ln^2)} =
 - \frac{i\alpha m}{8\pi D} Y(\sqrt{s}, 0, 0) \ln^2 \frac{Q^2}{m^2},
\ee
The following properties of these double-logarithmic contributions are notable: it is produced by the intermediate states with the maximal kinematically possible invariant mass, $W = \sqrt{s}$. Therefore the elastic contribution does not take part here, and energy dependence of asymmetry has resonance form with maxima at the positions of prominent resonances.

The double-logarithmic contribution has the following asymptotic at small $Q^2$:
\begin{equation} \label{atlowQ}
 B_n^{(\ln^2)} \sim Q^3 \ln^2{\frac{Q^2}{m^2}}\;.
\end{equation}
in agreement with the results of Ref.~\cite{Gorchtein}.

Another approach was used in Ref.~\cite{AfanMerCorr}.
While in Ref.~~\cite{ourBNSA} we searched for the terms with the highest power of the large logarithm,
the authors of Ref.~\cite{AfanMerCorr} seek for the slowest-decreasing terms at $Q\to 0$.
The result was the following:
\begin{equation} \label{AfBn}
 B_n \approx B_n^{(\ln)} = - \frac{2 m (s-M^2)^2}{\pi^2 D} \left(G_E+\tfrac{Q^2}{4M^2} G_M \right) 
 Q \ln \frac{Q^2}{m^2}\ \sigma_{\rm tot},
\end{equation}
where $\sigma_{\rm tot}$ is total photoabsorption cross-section on the proton, i.e.
the cross-section of the reaction $\rm \gamma p \to X$.

Note that Eqs.(\ref{AfBn}) and (\ref{Bn}) are distinct contributions and should in general be added together. The approximation of Eq.(\ref{AfBn}) is valid for very small scattering angles,
\begin{equation}
 \sin^2 \frac{\theta}{2} \, \ln \frac{Q^2}{m^2} \ll 1,
\end{equation}
while the double-logarithmic approximation of Eq.(\ref{Bn}) is valid upon the reverse condition
\begin{equation}
 \sin^2 \frac{\theta}{2} \, \ln \frac{Q^2}{m^2} \gg 1.
\end{equation}
%

%%%==================================================================================
\section{Calculation of the TPE amplitude}\label{sec:TPEcalc}

The calculation of real part of the TPE amplitude is more complicated task, since the intermediate hadronic state is here virtual and,
regardless of the total energy, states of all masses can contribute.
There are two main ways of such calculation: ``hadronic'' approach (Sec.\ref{sec:hadronic}) and ``parton-quark'' one, suitable for high $Q^2$ (Sec.\ref{sec:highQ2}).

\subsection{Hadronic approach}\label{sec:hadronic}

\begin{figure}
 \centering
 $\vcntr{\includegraphics[width=0.125\textwidth]{m2c.eps}}
	\quad\approx\quad \vcntr{\includegraphics[width=0.125\textwidth]{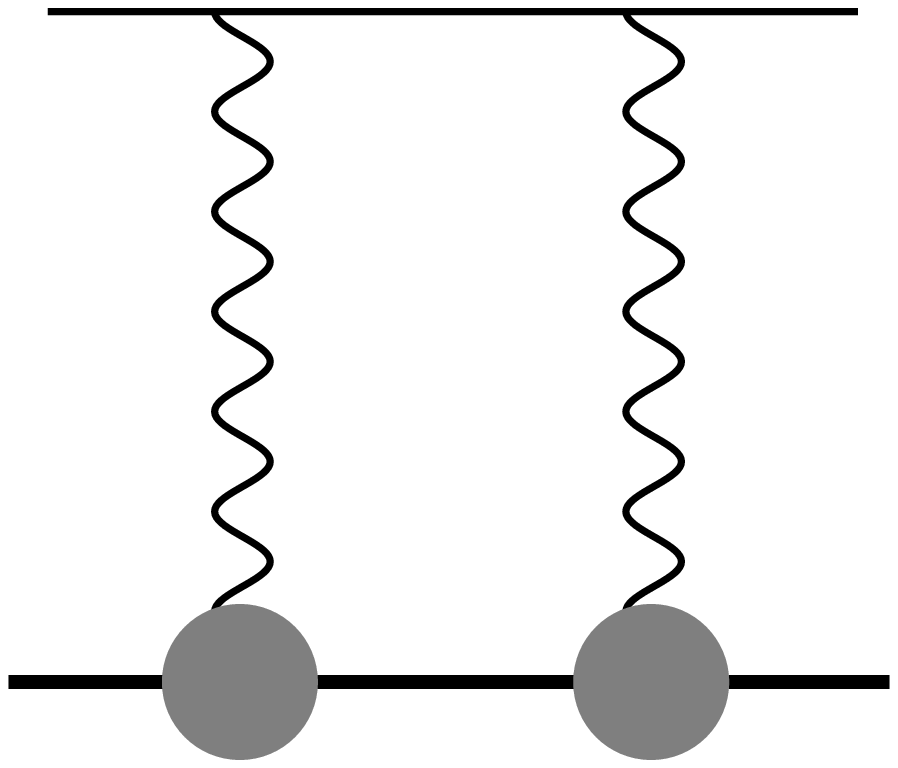}}
	\quad+\quad \vcntr{\includegraphics[width=0.125\textwidth]{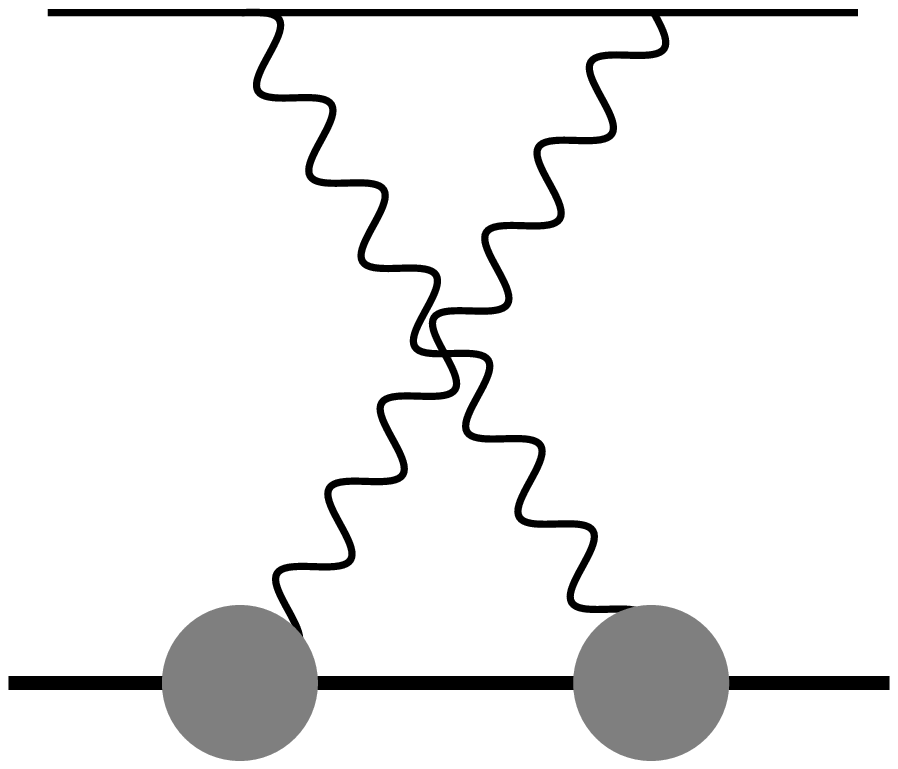}}$
 \caption{The elastic contribution} \label{diag-elas}
\end{figure}

The ``hadronic'' approach implies that the intermediate states in the lower part of diagram Fig.~\ref{diag2}d are sets of hadrons.
Certainly, unlimited number of variants is possible here, and just lowest-mass states are usually taken into account.

As with calculation of imaginary part, we distinguish ``elastic contribution'', which arises when the intermediate state is the proton itself, and inelastic one, coming from all other intermediate states.
The elastic contribution is easier to calculate, and in many situations it is the dominant one.
Also, the IR-divergent terms in the amplitude come exclusively from the elastic contribution; the inelastic one is always IR-finite.
Only the elastic contribution is to be considered for $e\mu$ (instead of $ep$) scattering, which was done in Ref.~\cite{mue-mue}.

The difference of the $ep$ scattering from the $e\mu$ case is that the proton has internal structure, and thus the question arises,
what is its propagator and $pp\gamma$ vertex function? It was argued that, since one of the protons entering the vertex,
is virtual, it should differ from the usual $pp\gamma$ e.m. vertex and contain so-called off-shell FFs,
which are not known experimentally.

Nevertheless, as an approximation, the TPE diagram was calculated in Ref.~\cite{bmt0,bmt} under assumption that
the proton propagator is the same as that of point Dirac particle and the $pp\gamma$ vertex coincide with the usual on-shell one:
\be
 \Gamma_\mu(q) = F_1(q^2) \gamma_\mu - \frac{1}{4M} F_2(q^2) [\gamma_\mu,\hat q],
\ee
This expression was chosen because it preserves gauge invariance, while other common one, equivalent in the case of on-shell proton,
\be
 \Gamma_\mu(q) = (F_1 + F_2) \gamma_\mu - F_2 \frac{(p+p')_\mu}{2M}
\ee
does not.
This approximation is schematically depicted in Fig.\ref{diag-elas}: the full TPE contribution is approximated by the sum of two diagrams, called ``box'' and ``crossed-box'' diagrams. They are related by crossing symmetry, thus the scattering amplitude is
\be
 \CM = \CM^\text{(box)} + \CM^\text{(xbox)} = \CM^\text{(box)}(\nu,t) - \CM^\text{(box)}(-\nu,t)
\ee
Some technical details of the calculation are given below.

\subsubsection{Elastic contribution}\label{sec:elastic}

The expression, corresponding to the box diagram in Fig.~\ref{diag-elas}, is
%\begin{eqnarray}
%  \Im\CM = \frac{1}{8\pi^2} \int \frac{(4\pi\alpha)^2}{q_1^2 q_2^2}
%    \bar u' \gamma_\mu (\hat k'' + m) \gamma_\nu u \cdot
%    \bar U' \Gamma_\mu(q_2) (\hat p''+M) \Gamma_\nu(q_1) U \times \nonumber \\
%  \times \theta(k''_0)\delta(k''^2-m^2) \theta(p''_0)\delta(p''^2-M^2) d^4k''.
%\end{eqnarray}
%
\be \label{ElasticContr}
 i \CM^\text{(box)} = \left (\frac{\alpha}{\pi}\right)^2 \int  d^4 k'' \frac{\bar u' \gamma_\mu (\hat k''+m) \gamma_\nu u \cdot
 \bar U' \Gamma_\mu(q_2) (\hat p''+M) \Gamma_\nu(q_1) U}{
 q_1^2 q_2^2 (k''^2-m^2)(p''^2-M^2)}
\ee
First, the $\gamma$-matrix structure of the formula must be reduced to that of Eq.(\ref{genAmp}).
This can be done in two stages.
It is well-known that any product of $\gamma$ matrices can be represented as a linear combination of the 16 structures:
$1$, $\gamma_\mu$, $[\gamma_\mu, \gamma_\nu]$, $\gamma_5\gamma_\mu$ and $\gamma_5$.
Then we use the fact, that in Eq.(\ref{ElasticContr}) those matrices are sandwiched between on-shell particle spinors.
It can be shown that
\bea
\tfrac{1}{2} q^2 \, \bar u' \gamma_5 \gamma_\mu u & = & -m q_\mu \, \bar u' \gamma_5 u - i\epsilon_{\mu\nu\sigma\tau} K^\sigma q^\tau \, \bar u' \gamma^\nu u, \\
\tfrac{1}{4} q^2 \, \bar u' [\gamma_\mu, \gamma_\nu] u & = & - i \epsilon_{\mu\nu\sigma\tau} K^\sigma q^\tau \, \bar u' \gamma_5 u
    + m \left( q_\mu \bar u' \gamma_\nu u - q_\nu \bar u' \gamma_\mu u \right)
    + \bar u'u \left( K_\mu q_\nu - K_\nu q_\mu \right).
\eea
(and similarly for the proton spinors, with the replacement $K_\mu \to P_\mu$, $q_\mu \to -q_\mu$, $m \to M$).
Using these identities, $\cal M$ can be represented as a linear combination of the four structures, of the form
\be
 \bar u' \gamma_\mu u \; \bar U' \gamma_\nu U, \quad
 \bar u' \gamma_5 u \; \bar U' \gamma_\mu U, \quad
 \bar u' \gamma_\mu u \; \bar U' \gamma_5 U, \quad \text{and} \quad
 \bar u' \gamma_5 u \; \bar U' \gamma_5 U.
\ee
The 2nd and 3rd combinations would violate $T$-invariance and thus do not actually appear;
and the last combination does not appear due to negligibly small electron mass.

Now the task is to calculate the coefficients in front of these structures and compare the resulting amplitude with Eq.(\ref{genAmp})
to obtain generalized FFs.
The combinations that would violate $T$ invariance automatically receive zero coefficients after the integration.

At this stage, we obtain for the generalized FFs expressions like
\be \label{FFintegral}
   \delta\CG = \sum_{i,j=1}^2 \int d^4 k''\frac{F_i(t_1)F_j(t_2)}{t_1 t_2}
   \left[
     \frac{A_{ij}(t_1,t_2,\nu,t)}{(k''^2-m^2)(p''^2-M^2)}
   + \frac{A_{k,ij}(t_1,t_2,\nu,t)}{k''^2-m^2}
   + \frac{A_{p,ij}(t_1,t_2,\nu,t)}{p''^2-M^2}
   + A_{1,ij}(t_1,t_2,\nu,t)
  \right]
\ee
which contain elastic proton FFs $F_i$. The functions $A_{ij}$ are polynomials in $t_1$, $t_2$ and rational functions in $\nu$. To proceed further, the FFs are parameterized as a sum of fixed poles:
\be
  \frac{F_i(t)}{t} = \sum_a \frac{c_{ia}}{t-m_a^2},
\ee
The integral (\ref{FFintegral}) is then represented as a linear combination
\bea \label{NPintegral}
 \delta \CG &=& \sum_{ijab} c_{ia} c_{jb}
   \int \frac{d^4 k''}{(t_1-m_a^2)(t_2-m_b^2)} \times \\
  && \times \left\{ 
     \frac{A_{ij}(m_a^2,m_b^2,\nu,t)}{(k''^2-m^2)(p''^2-M^2)}
   + \frac{A_{k,ij}(m_a^2,m_b^2,\nu,t)}{k''^2-m^2}
   + \frac{A_{p,ij}(m_a^2,m_b^2,\nu,t)}{p''^2-M^2}
   + A_{1,ij}(m_a^2,m_b^2,\nu,t)
 \right\}   \nonumber
\eea
The integrals of the form
\bea
&& \int \frac{d^4 k''}{(t_1-m_a^2)(t_2-m_b^2)(k''^2-m^2)(p''^2-M^2)}, \\
&& \int \frac{d^4 k''}{(t_1-m_a^2)(t_2-m_b^2)(k''^2-m^2)}, \\
&& \int \frac{d^4 k''}{(t_1-m_a^2)(t_2-m_b^2)}, \\
\eea
entering Eq.(\ref{NPintegral}) are known as 4-point, 3-point and 2-point functions, respectively \cite{tHooft}, and can be calculated either analytically or numerically.

Finally the IR-divergent contribution should be subtracted (see also discussion in Sec.~\ref{sec:IR}), which is equal
\be
\delta\CG_E^\text{(IR)} = f^\text{(IR)} G_E, \qquad
\delta\CG_M^\text{(IR)} = f^\text{(IR)} G_M, \qquad
\delta\CG_3^\text{(IR)} = 0
\ee
where the factor $f^\text{(IR)}$ depends on the particular method of radiative corrections calculation. In the prescription of Mo-Tsai \cite{MoTsai}, which was widely used in experiments, it is equal
\be
 f^\text{(Mo-Tsai)} = \frac{\alpha}{\pi} 
   \left\{   \ln\frac{4 M^2 \lambda^2}{\nu^2-t^2}\ln\frac{\nu-t}{\nu+t}
      - \mathop{\rm Li_2}\left(1-\frac{\nu-t}{2M^2}\right)
      + \mathop{\rm Li_2}\left(1-\frac{\nu+t}{2M^2}\right)
   \right\}
\ee
In some recent experiments the so-called Maximon-Tjon prescription \cite{MaximonTjon} is used instead of Mo-Tsai, which (naturally) has the same divergent part, but differs from the latter in finite terms:
\be
 f^\text{(MT)} = \frac{\alpha}{\pi} \ln\frac{\lambda^2}{-t}\ln\frac{\nu-t}{\nu+t}
\ee

\subsubsection{Time-like region problem}
However, there is a problem with this approach, which was pointed out in Ref.~\cite{ourBox}.
The procedure, described above, involves FF fitting.
But the FFs are measured experimentally only in the space-like region ($Q^2 > 0$),
and naturally the fit is done over that region only.
In the time-like region ($Q^2 < 0$) the difference between the actual FFs and the fit can be significant.
But the integration in Eqs.(\ref{ElasticContr}, \ref{FFintegral}) goes over both space-like and time-like regions, and therefore the results of calculations using such fit are doubtful.

To overcome this problem, in Ref.~\cite{ourBox} another method of integration was used.
Using the analytic properties of FFs, Eq.(\ref{FFintegral}) was transformed via Wick rotation, resulting in the other integral,
containing FFs for space-like $Q^2$ only:
\be
 \delta\CG = \sum_{i,j=1}^2 \int_{t_1,t_2\le 0} {\cal K}_{ij}(t_1,t_2) F_i(t_1) F_j(t_2) dt_1 dt_2,
\ee
where ${\cal K}_{ij}(t_1,t_2)$ are certain known functions.

This result implies that fitting FF in the space-like region only is fine,
as long as the analytic structure of FFs is preserved by the fit,
i.e. all singularities lie on the negative real axis.
In this case the integral effectively depends on FFs at $Q^2 > 0$ only.

Trying calculation with different FF parameterizations, it was found that results are almost insensitive to it.

\subsubsection{Dispersion approach}

However, the problem of the off-shell FFs remained unsolved. To resolve this problem, the dispersion method was proposed \cite{ourDisp}.

The idea is that at first, the imaginary part of the TPE amplitude is calculated.
This is done with the help of the unitarity condition (\ref{unit1}), where the intermediate states are on-shell,
therefore one can employ usual FFs, measured experimentally.
%For the elastic contribution, the corresponding equation differs from Eq.~(\ref{ElasticContr}) by replacement
%\be
%\frac{1}{(k''^2-m^2)(p''^2-M^2)} \to 2\pi^2 \theta(k''_0)\delta(k''^2-m^2) \theta(p''_0)\delta(p''^2-M^2)
%\ee

Then the real part of the amplitude is reconstructed via the dispersion relations, and it appears that
(in the case of single-particle intermediate state, either elastic or inelastic)
such reconstruction can be done analytically and independently of particular FFs.

Calculations with this method were first done for the elastic intermediate state \cite{ourDisp}.
It was found that the elastic contribution, calculated in Ref.~\cite{ourDisp} with dispersion approach, differs from the results of Refs.~\cite{bmt,ourBox}, but the difference is very small numerically.

\subsubsection{Inelastic contribution}

Later on, calculations like \cite{bmt} were performed for the inelastic contribution: namely, $\Delta(1232)$ \cite{bmtDelta}
and higher resonances \cite{bmtRes} as intermediate states.
The width of resonances was neglected: they were considered as zero-width particles with proper spin-parity and on-shell transition FFs.
For their propagators, usual propagators of spin-1/2 and spin-3/2 particles were used.

The results were that contributions of the resonances are smaller than the elastic one, the largest of them is the $\Delta(1232)$ contribution,
and the contributions of different resonances have different signs, partly cancelling each other.

The approach of Ref.~\cite{ourBox} can also be applied in this case, showing that it is still sufficient
to fit the transition amplitudes at $Q^2>0$ only \cite{ourBoxRes}.

However, it is clear that neglecting the resonance width completely is not good idea,
as the width may be even comparable to mass (as in R\"oper resonance $N^*(1440)$); also the problem of proper choice of propagator and transition FFs remains unsolved.
Thus, the dispersion approach of Ref.~\cite{ourDisp} was applied to the calculation of the inelastic contribution.
In Ref.~\cite{ourDelta}, the contribution of zero-width $\Delta$ resonance was calculated within this approach and in Ref.~\cite{ourP33}
the full $P_{33}$ channel of the $\pi N$ system was included, effectively taking into account $\Delta$ resonance with realistic width and shape
(as $\Delta$ almost 100\% consists of $\pi N$).
In Ref.~\cite{ourPiN} the same approach was used to include all $\pi N$ states with spin 1/2 and 3/2 (namely, $S_{11}$, $S_{31}$, $P_{11}$, $P_{31}$, $P_{13}$, $P_{33}$, $D_{13}$ and $D_{33}$ channels). Here the $\pi N$ system with c.m. energy $W$ was treated as a single particle with proper spin-parity, mass $W$ and transition FFs derived from the pion photoproduction amplitudes, which, in turn, were taken from the MAID model \cite{MAID}.

The results were the following.
At small $Q^2$ the inelastic contribution to TPE amplitudes is very small (negligible compared to the elastic one);
at intermediate energies it has resonance shape with maxima near the positions of prominent resonances, which are sharp if we neglect resonance width (Ref.~\cite{ourDelta})
and become smooth if we properly account it (Refs.~\cite{ourP33, ourPiN});
and at high $Q^2$ it is gradually growing with $Q^2$.

The main contribution among all inelastic channels comes from the $P_{33}$ channel where $\Delta(1232)$ lives, and overall, larger contributions come from the channels with quantum number of lightest resonances.

Though the TPE corrections to the cross-section and to the magnetic FF are dominated by the elastic contribution, it was found that the agreement with experimental data improves after taking into account the inelastic contribution.

It was also found that the correction to the $G_E$ FF due to inelastic states is relatively large, growing almost linearly with $Q^2$ at
$Q^2 \gtrsim 2 \GeV^2$, and soon exceeds the elastic one.

\subsection{High $Q^2$}\label{sec:highQ2}

\begin{figure}%[b]
\def\www{0.13}
 \centering
 (a)
 \includegraphics[width=\www\textwidth]{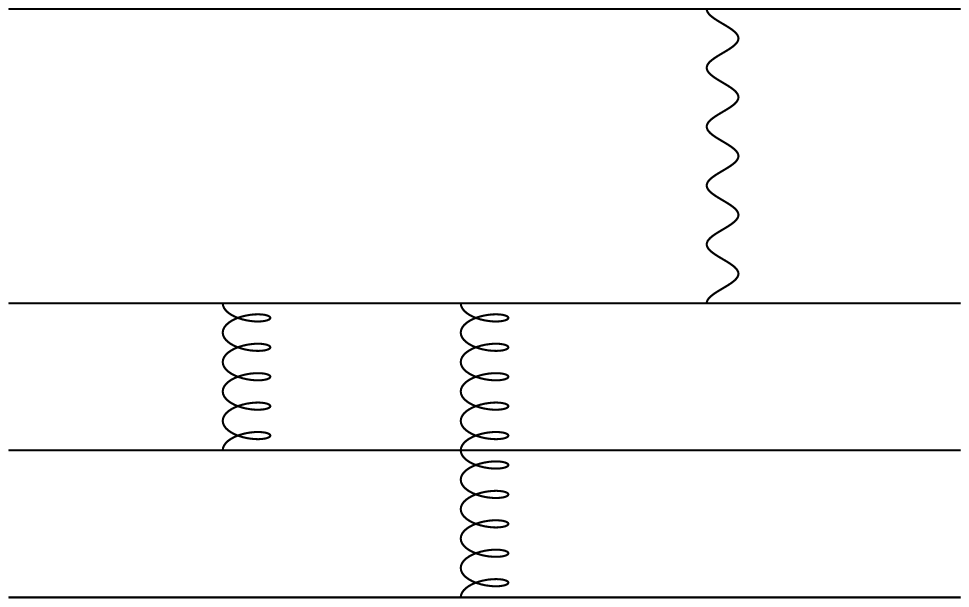}\hfil
 \includegraphics[width=\www\textwidth]{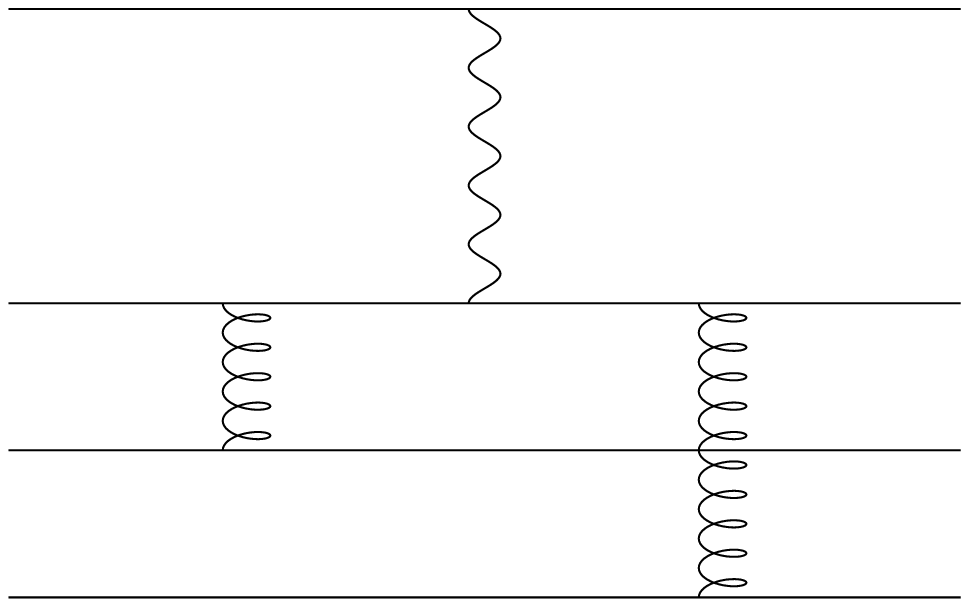}\hfil
 \includegraphics[width=\www\textwidth]{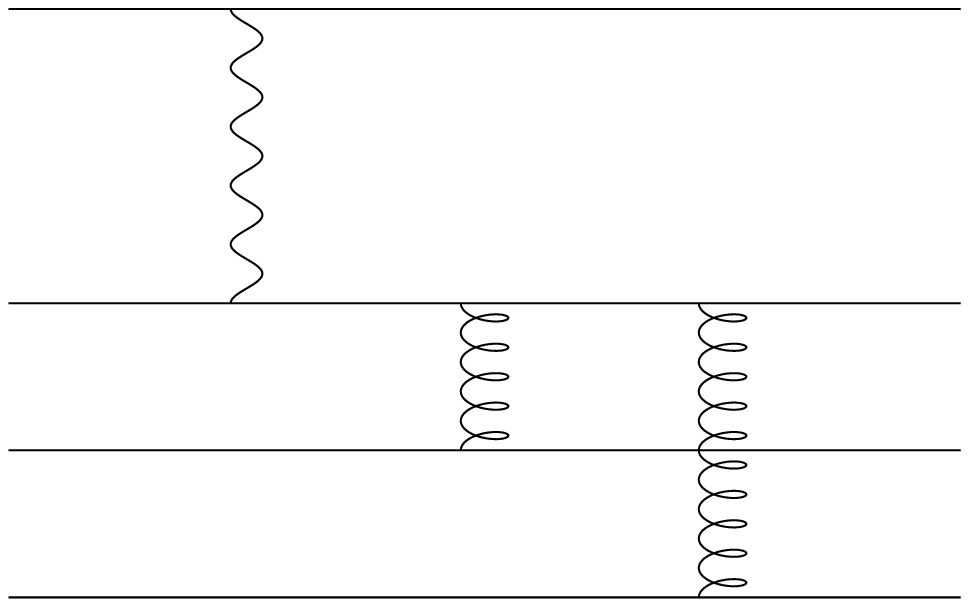}\hfil
 \includegraphics[width=\www\textwidth]{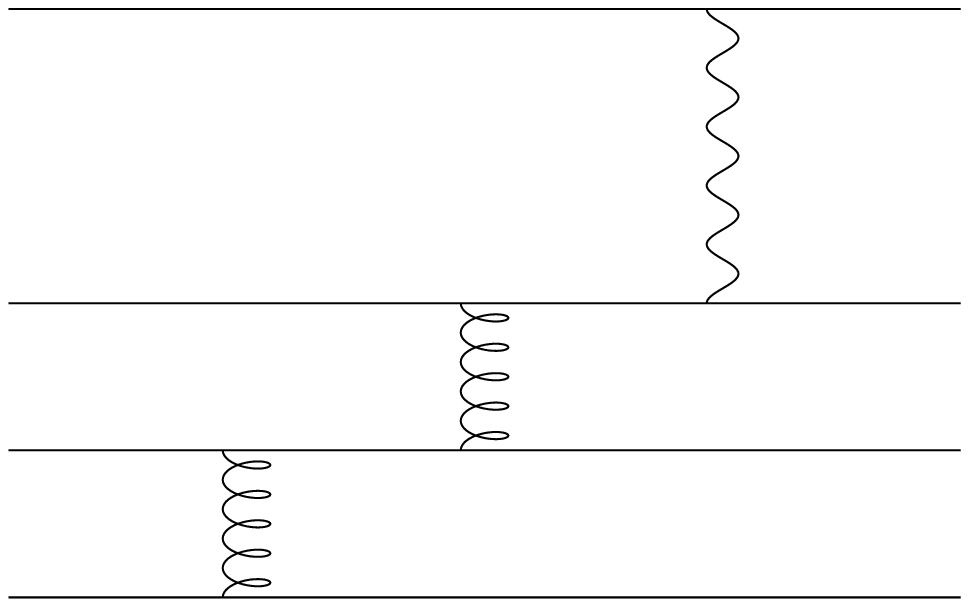}\hfil
 \includegraphics[width=\www\textwidth]{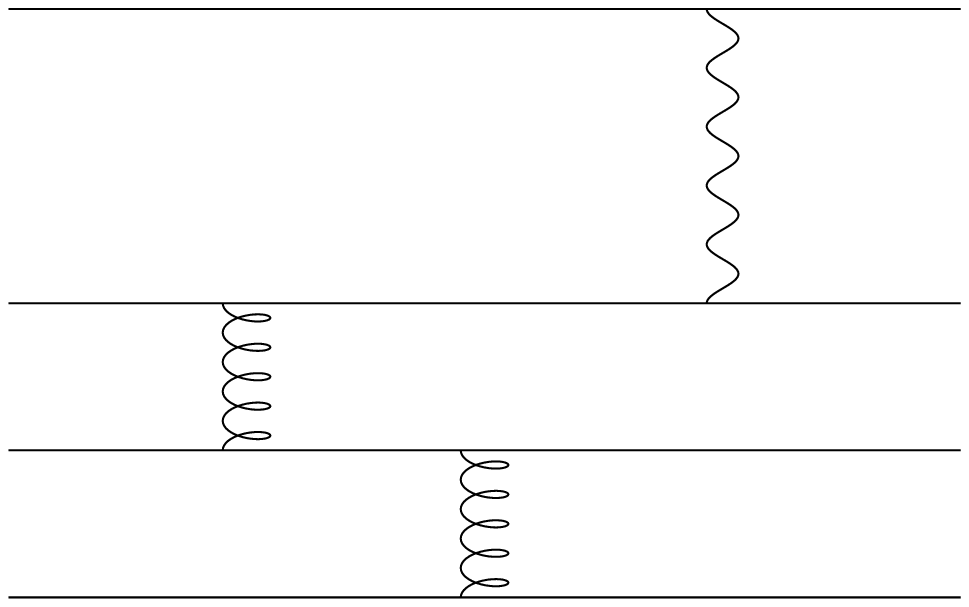}\hfil
 \includegraphics[width=\www\textwidth]{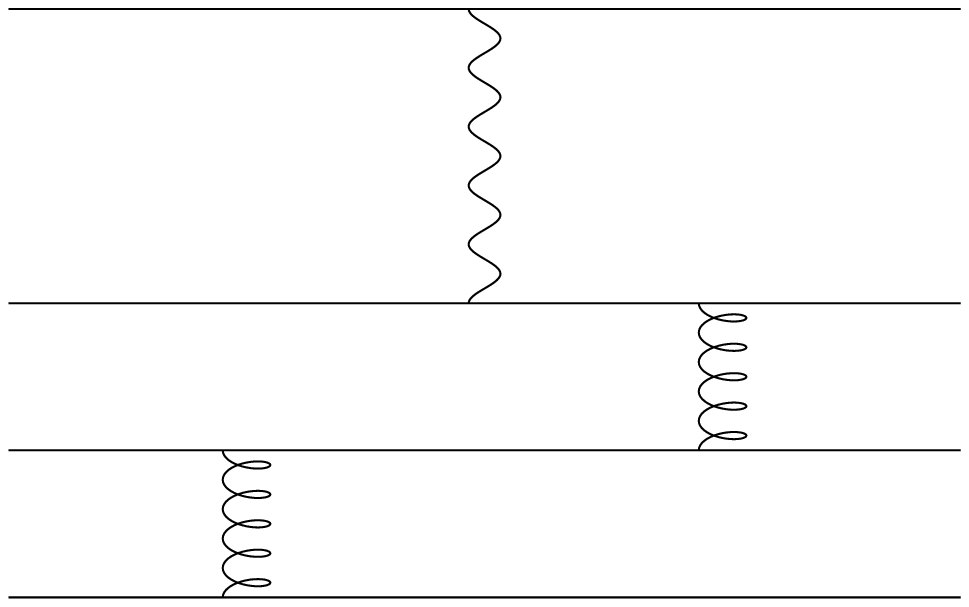}\hfil
 \includegraphics[width=\www\textwidth]{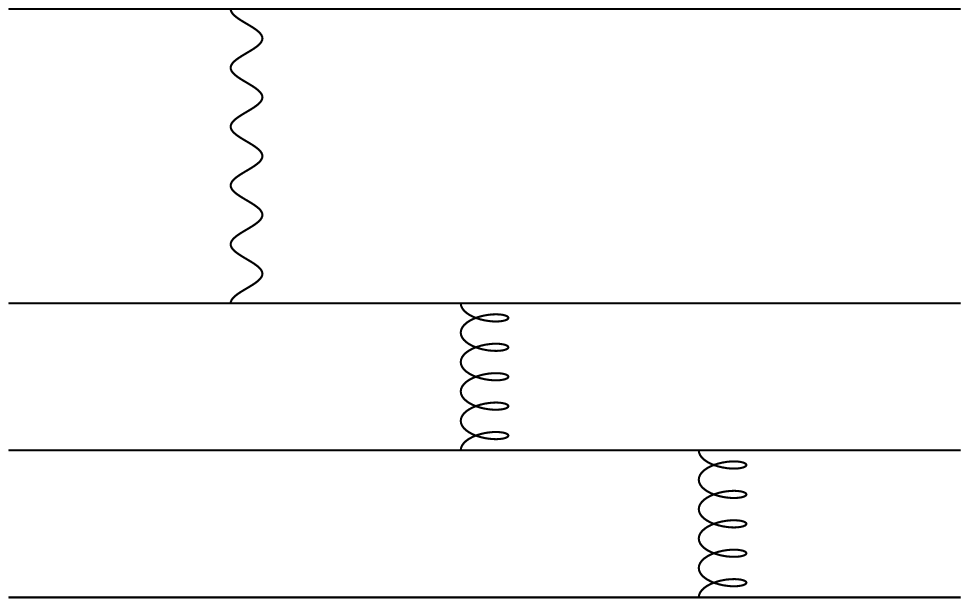}\\[5mm]
 (b)
 \includegraphics[width=\www\textwidth]{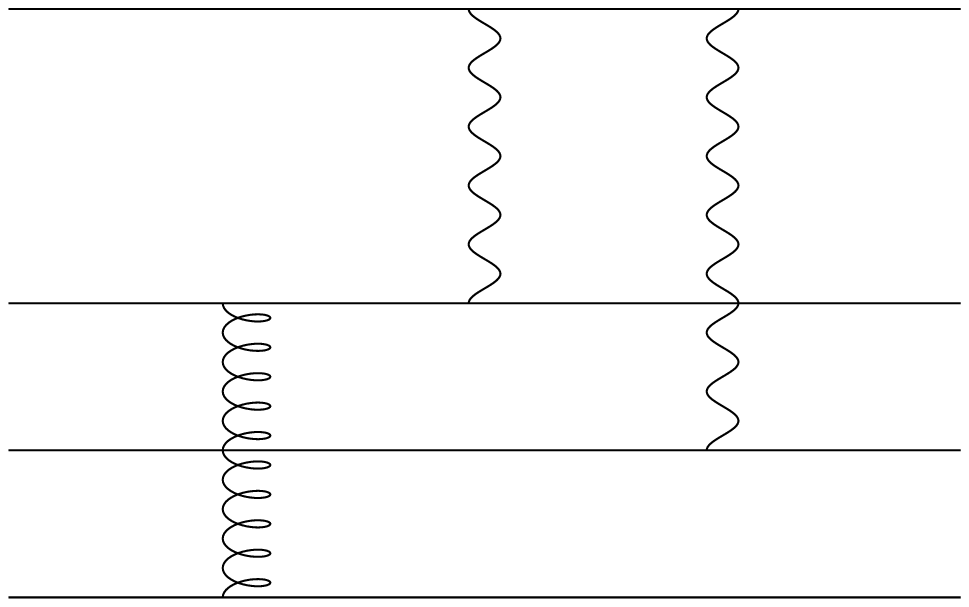}\hfil
 \includegraphics[width=\www\textwidth]{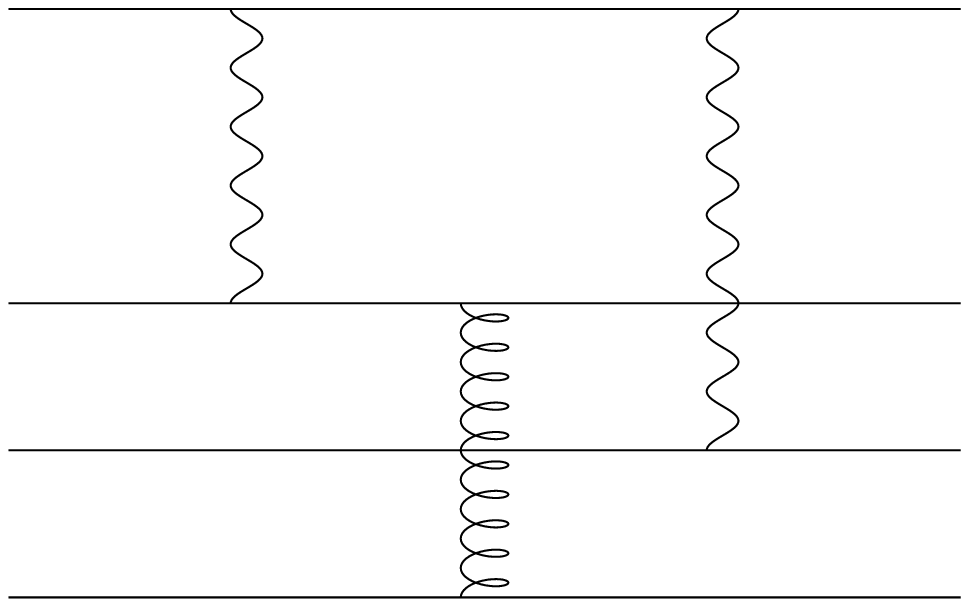}\hfil
 \includegraphics[width=\www\textwidth]{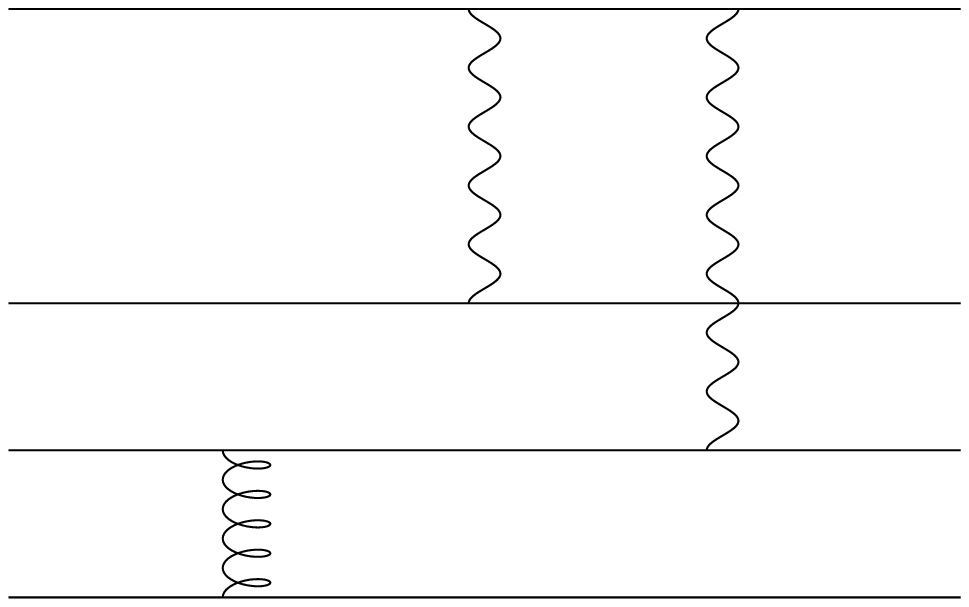}\hfil  
 \includegraphics[width=\www\textwidth]{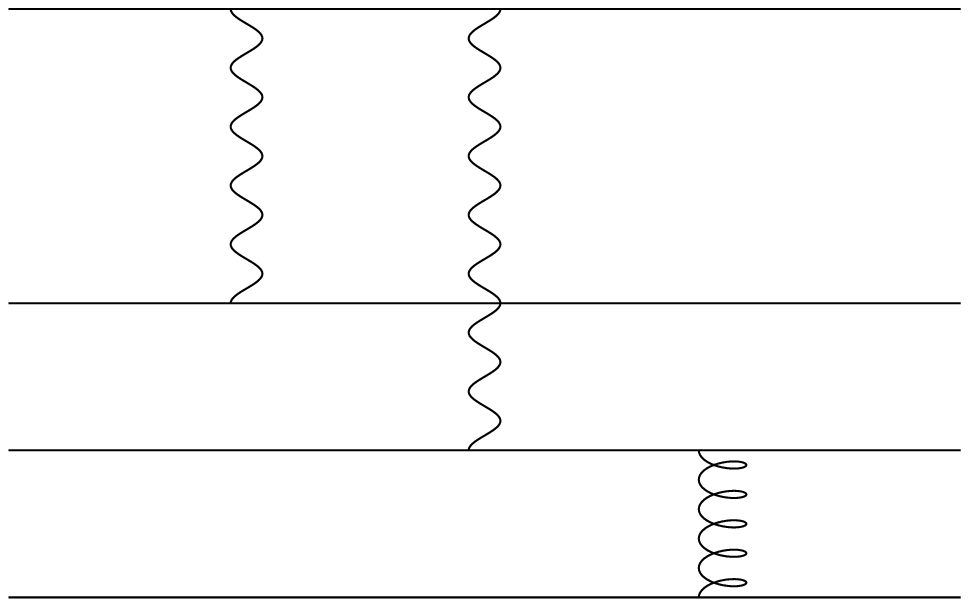}\hfil
 (c)
 \includegraphics[width=\www\textwidth]{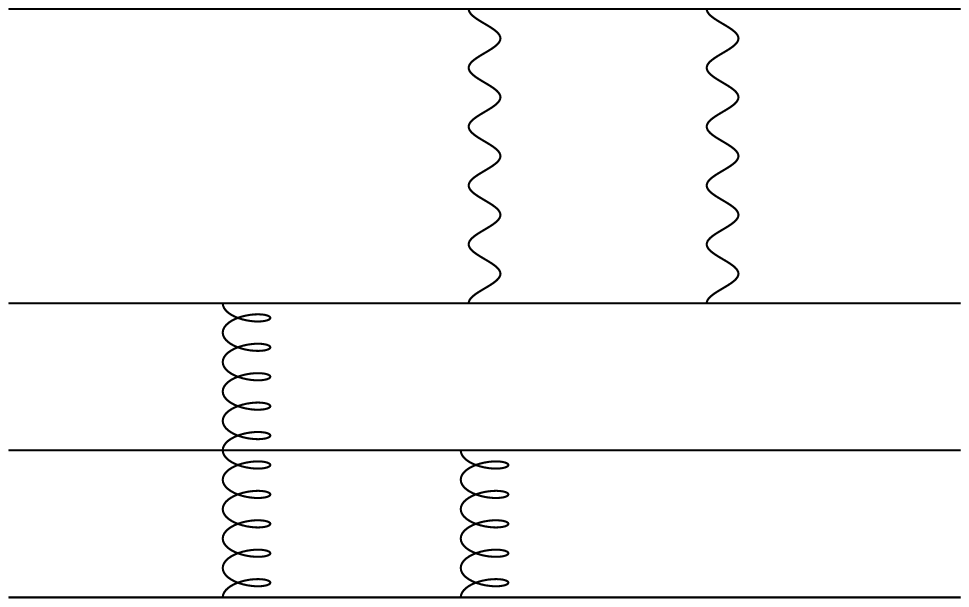}\hfil
 (d)
 \includegraphics[width=\www\textwidth]{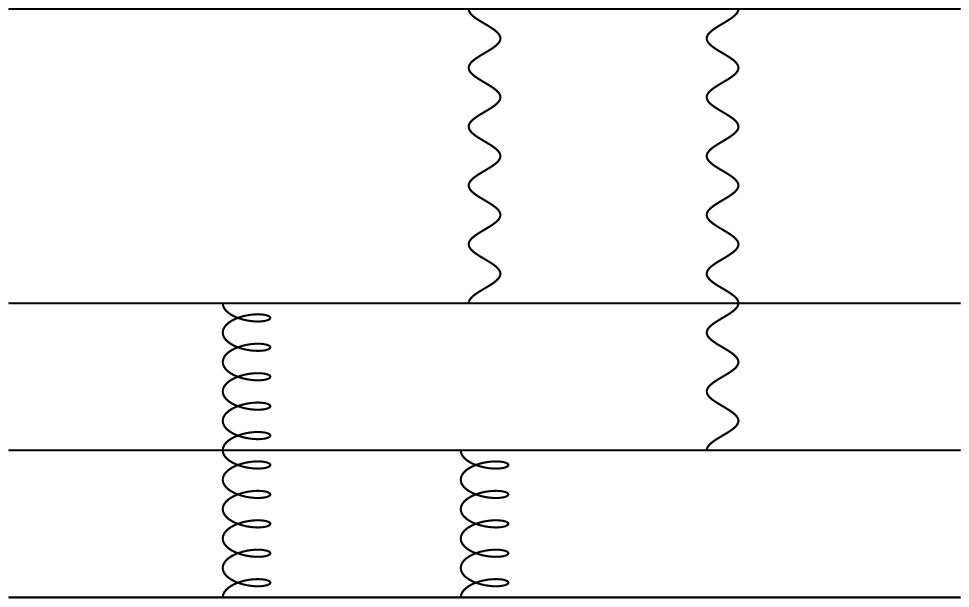}
 \caption{pQCD diagrams for $eN\to eN$: OPE (a),
 TPE, leading order (b), subleading order~(c,d).}
 \label{QCD-diagr}
\end{figure}
At high $Q^2$, the hadronic approach becomes doubtful, because of large number of intermediate states involved.
On the other hand, at high $Q^2$ proton-virtual photon interaction can be treated with the help of special technics,
developed for that case: generalized parton distributions (GPDs) and perturbative quantum chromodynamics (pQCD).

The GPD is a generalization of parton distribution functions, which are used in describing deep inelastic scattering \cite{GPD-rev}.
In the GPD model of Ref.~\cite{GPD}, both virtual photons interact with the same quark (on contrary, in pQCD such diagrams are suppressed by a factor $\alpha_s$, since they need extra gluon to be exchanged between quarks, see below).
The TPE amplitude is obtained as a convolution of the TPE amplitude of the elastic electron-quark process with the distribution of quarks in proton --- GPD.
Authors of Ref.~\cite{GPD} claim that the diagrams, in which two photons interact with different quarks, are ``subleading in $Q^2$ because of the momentum mismatches in the wavefunctions''.

The pQCD can be used, in particular, to describe high-$Q^2$, high-energy exclusive reactions with hadrons, such as elastic lepton scattering \cite{QCD1}.
Though straightforward application of pQCD yields results, which are not very consistent with the experiments
(in particular, magnetic FF of the proton becomes zero and that of the neutron turns positive),
it was shown that introducing phenomenological quark distribution amplitude allows to obtain reasonable agreement at currently accessible experimental energies. In pQCD approach, the scattering amplitude is split into hard ``core'', which describes scattering on a set of (asymptotically free) quarks, and calculated according to QCD perturbation theory,
and quark distribution amplitudes, which connect physical hadron with multi-quark states and are determined empirically
with the help of various sum rules.

PQCD representation of TPE has an important feature: comparing pQCD diagrams for OPE (Fig.~\ref{QCD-diagr}a) and TPE (Fig.~\ref{QCD-diagr}b), one can see that the former includes 1 photon and 2 gluon exchanges, producing a factor of $\alpha\alpha_s^2$, whereas for the TPE we have 2 photons and 1 gluon, which gives a factor of $\alpha^2\alpha_s$. This way in pQCD the ratio TPE/OPE is of order $\alpha/\alpha_s$, which is much larger than naive $\alpha$, and raises with $Q^2$, as $\alpha_s$ decreases.

In Ref.~\cite{ourQCD} the TPE corrections to the elastic $eN$ scattering were calculated in leading-order pQCD.
In this approximation, only non-spin-flipping amplitudes exist — $\delta \CG_M$ and $\delta \CG_3$, whereas $\delta\CG_E$ (as $G_E$ itself) cannot be assessed: they are subleading-order effects. It is convenient to use normalized TPE amplitudes: $\delta \CG_M/G_M$ and $\delta\CG_3/G_M$, where both numerator and denominator are calculated according to pQCD. This way we also avoid uncertainty coming with the absolute normalization of the quark distribution amplitudes.

The results are:
\begin{equation} \label{ratio}
 \left( \frac{\delta G_M}{G_M}, \frac{\delta\CG_3}{G_M} \right) =
   -\frac{3\alpha}{\alpha_s}
    \frac{\<\phi(y_i)|(T_{\delta G_M},T_{\delta\CG_3})|\phi(x_i)\>}
    {\<\phi(y_i)|T_{G_M}|\phi(x_i)\>}
\end{equation}
with
\bea
 T_{G_M} & = & (1 + h_1 h_3) \left\{
     \frac{2e_1}{x_3 y_3 (1-x_1)^2 (1-y_1)^2} + 
     \frac{2e_1}{x_2 y_2 (1-x_1)^2 (1-y_1)^2} + 
     \right. \nonumber \\ && \left.     
     \frac{e_2}{x_1 y_1 x_3 y_3 (1-x_1) (1-y_3)} -
     \frac{e_1}{x_2 y_2 x_3 y_3 (1-x_1) (1-y_3)} -
     \frac{e_1}{x_2 y_2 x_3 y_3 (1-x_3) (1-y_1)}
  \right\} \\
 T_{\delta G_M} & = &
   \frac{e_1 e_2 (1 - h_1 h_3)}{x_2 y_2 x_3 y_3 (1-x_2)(1-y_2)}
   \frac{ (\nu-q^2)/(1-x_2) + (\nu+q^2)/(1-y_2) - 2\nu }
   {\nu(x_2-y_2)-q^2(x_2+y_2-2 x_2 y_2)+i0} \\
 T_{\delta\CG_3} & = &
   \frac{e_1 e_2 (1 - h_1 h_3)}{x_2 y_2 x_3 y_3 (1-x_2)(1-y_2)}
   \frac{2\nu}{\nu(x_2-y_2)-q^2(x_2+y_2-2 x_2 y_2)+i0}
\eea
where $\phi(x_i) \equiv \phi(x_1, x_2, x_3)$ are quark distribution amplitudes, $e_i$ are quark charges, and $h_i$ are doubled helicities.
Soon afterwards the same results were obtained in Ref.~\cite{KivelQCD}.

Main features of this result are:
\begin{itemize}
\item the TPE amplitude is proportional to $\alpha/\alpha_s$, which means approximately logarithmic growing with $Q^2$,
\item it has identical $\ve$-dependence at any fixed $Q^2$,
\item absence of IR divergence — it is easy to see that IR-divergent terms are subleading in $\alpha_s$,
\item the main contribution in pQCD regime comes from the area where both virtual photons are hard, $Q_1^2 \sim Q_2^2 \sim Q^2$.
\end{itemize}

\begin{figure}
\centering
\includegraphics[width=0.45\textwidth]{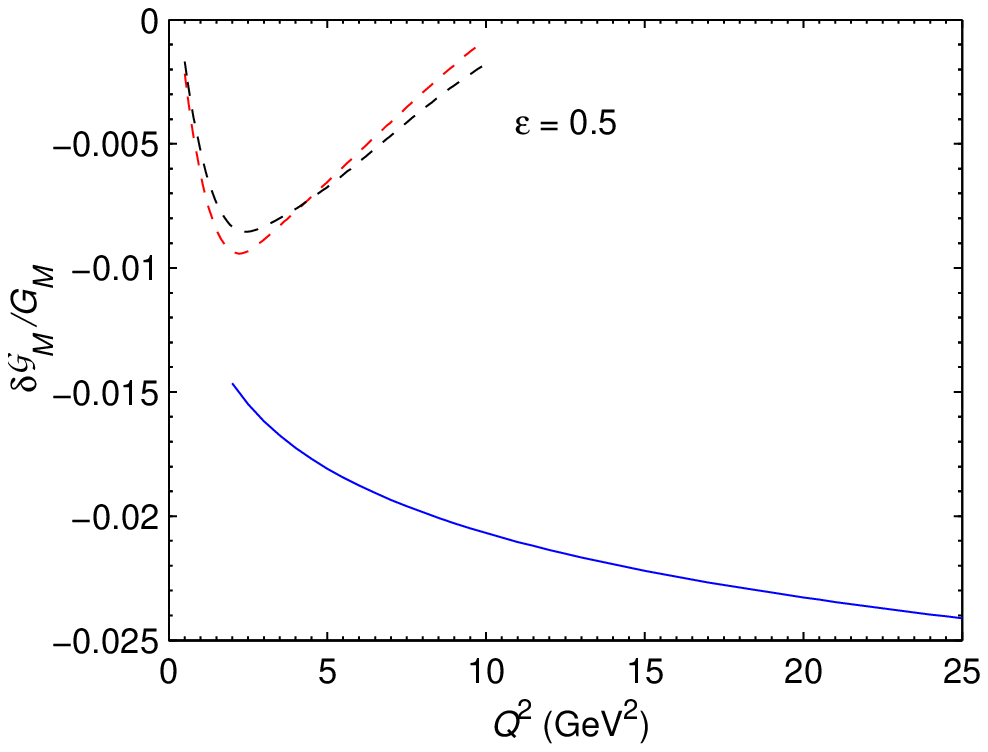}
\includegraphics[width=0.45\textwidth]{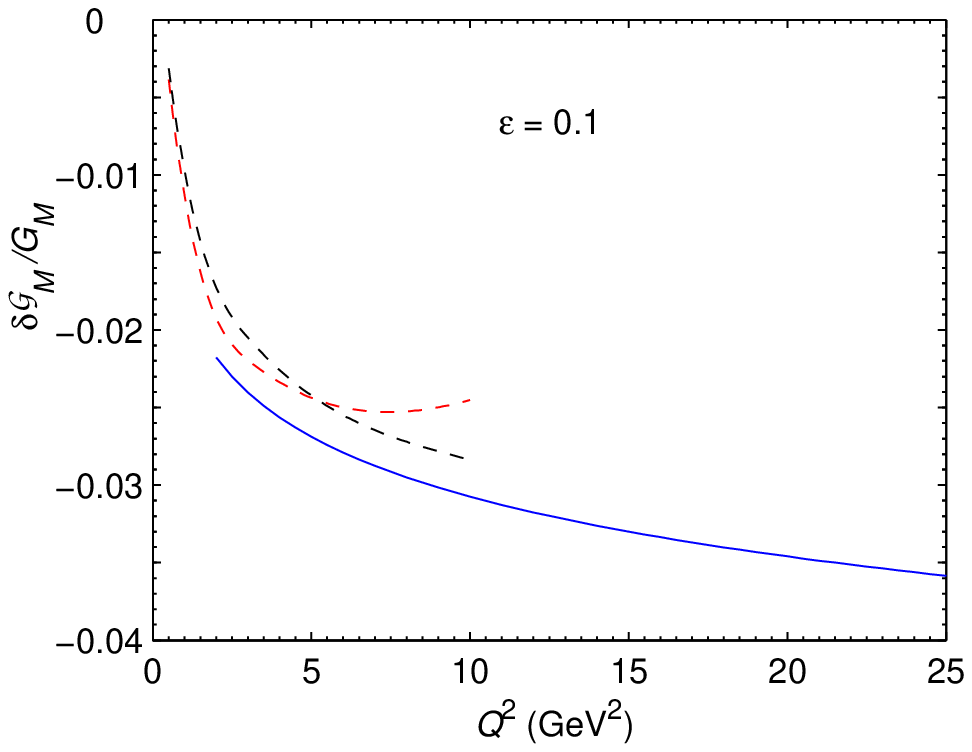}
\caption{TPE amplitude $\delta\CG_M$ vs. $Q^2$ at $\ve=0.5$ (left)
 and $\ve = 0.1$ (right). Dashed curves show ``hadronic approach'' calculations,
 with two different FF parameterizations.}\label{qcdfig}
\end{figure}

It was also shown that the results of ``hadronic'' and pQCD approaches are compatible with each other at lower $Q^2$, and rather smooth transition from ``hadronic'' to pQCD regime seems to occur at $Q^2 \gtrsim 3\GeV^2$, Fig.~\ref{qcdfig}.

In Ref.~\cite{SCET} authors argued that, though pQCD approach gives good approximation of ``hard contribution'', i.e. contribution where both virtual photons are hard, the ``soft'' contribution like one considered in Ref.~\cite{GPD} is not negligible at actual experiments kinematics, and should be added.
This idea was tried in Ref.~\cite{SCET}, where the soft contribution was calculated in so-called soft collinear exchange (SCET) approach.

\subsection{Low $Q^2$ and proton radius}
The behaviour of TPE amplitudes at low $Q^2$ and energies is of special interest not only by itself,
but also because this may affect proton radius extraction.

The proton electric radius $r_E$ is defined by
\be
 r_E^2 = -\frac{1}{6} \left. \frac{dG_E}{dQ^2} \right|_{Q^2=0}
\ee
and equals, in the nonrelativistic approximation, to the r.m.s. radius of the electric change distribution inside the proton.
This important quantity not only shows the proton ``size'', but influence other observables such as sizes of light nuclei, Lamb shift and hyperfine splitting in hydrogen, etc.
Similarly defined magnetic radius $r_M$ gives r.m.s. radius of the magnetic moment distribution
\be
 r_M^2 = -\frac{1}{6} \left. \frac{dG_M}{dQ^2} \right|_{Q^2=0}
\ee

The value of $r_E$ is usually determined from low-energy $ep$ scattering, by measuring $G_E$ form factor at reasonably small $Q^2$ and extrapolating to $Q^2 = 0$ to find the derivative.
This approach was however criticized in \cite{Sick}, since on the one hand, at larger $Q^2$ the result is influenced by higher-order (in $Q^2$) terms,
but on the other hand, too low-$Q^2$ measurements are more prone to systematic errors.

If $Q^2\to 0$ and $E\to 0$, the proton can be considered as a point particle and the TPE
reduces to second Born approximation for the scattering in Coulomb potential, which was studied long ago in Refs.~\cite{McKinley,Dalitz}.
The corresponding contribution to the cross-section is sometimes called ``Coulomb correction'':
\be
\frac{\delta\sigma}{\sigma} = \alpha\pi \frac{\sin \frac{\theta}{2}}{1+\sin \frac{\theta}{2}}
\ee
It was shown to affect proton radius extraction \cite{Rosenfelder}: after including Coulomb corrections, $r_E$ increased by $(0.008-0.013)$ fm depending on the fit strategy.

In a somewhat more precise approach, the proton is considered as a fixed source of external electric field
with the profile corresponding to its electric FF, thus the problem reduces to the relativistic electron scattering
in the external potential. The second Born approximation for this process was studied in Ref.~\cite{Lewis},
where the cross-section for general case and for some common potentials was obtained in analytical form.

From numerical results discussed in Sec.~\ref{sec:elastic}, the TPE amplitude changes its sign at $Q^2 \sim 0.3 \GeV^2$ and rather sharply grows at $Q^2\to 0$.
Note however, to prevent misunderstanding, that speaking of $Q^2\to 0$ we mean $Q^2 \ll M^2$ but still $Q^2 \gg m^2$,
otherwise our main approximation (ultrarelativistic electron) would not work.
For $Q^2 \ll m^2$ the results would be quite different and this is not discussed here.

In Ref.~\cite{ourLow} the low-$Q^2$ limit for the real part of the TPE amplitudes was obtained from the general relativistic formulae
via limiting procedure.
The results for $\CG_E$ at $Q^2\to 0$ are consistent with \cite{McKinley} (as at small energies electric FF dominates the cross-section), whereas for $\CG_M$ and $\CG_3$ they were new.

In Ref.~\cite{ourNR} the TPE amplitudes were calculated via the nonrelativistic approach (of course, nonrelativistic refers to proton and not electron),
which for the amplitude $\CG_E$ is equivalent to the results of \cite{Lewis};
it was found that at moderate $Q^2$ the real part of the amplitudes is described reasonably well,
but interestingly, the imaginary part strongly differs from the relativistic result, coinciding with it at $Q^2\to 0$ only.

The full TPE correction (not just the Coulomb one) was applied during extraction of proton radius in \cite{BlundenSick} and \cite{ourRadius}.
Ref.~\cite{BlundenSick} concludes that numerically, the difference between full TPE and just Coulomb correction is small (+0.0015 fm to $r_E$).
In Ref.~\cite{ourRadius} it was found that both electric and magnetic radii increase after taking TPE into account, $r_E$ by $\sim 0.01$ fm and $r_M$ by $\sim 0.03$ fm.

In the meantime, the proton electric radius was determined with the entirely different method:
measuring Lamb shift in muonic hydrogen \cite{rE-muon}.
Surprisingly, the obtained value of 0.841 fm was in striking disagreement with the previous results from the $ep$ scattering, 0.8768(69) fm \cite{rE-electron}.
This so-called ``proton radius puzzle'' gave rise to speculations whether there is some peculiarity in the muon electromagnetic interaction,
or the discrepancy is just the result of poor consistency of scattering approach.
To explore the former possibility, a muon-proton scattering experiments were proposed \cite{MUSE},
which, in turn, called for estimate of TPE effects in muon-proton scattering, discussed in Sec.\ref{sec:muon}.

On the other hand, large experiment on low-energy $ep$ scattering was performed at Mainz Microtron \cite{Bernauer}.
Its results were, in general, consistent with previous electron scattering experiments and still in disagreement with new muon hydrogen data.
In particular, the electric radius was found to be 0.879(8) fm.
Also, the authors found, that to achieve consistency with the polarization experiments some extra non-standard correction is required, which they cautiously interpret as ``TPE or other physics'' \cite{Bernauer2}.

%%%==================================================================================
\section{Experiments}\label{sec:experiment}

A number of experiments were performed to see the TPE effects caused by the real part of the amplitude directly.

The GEp-2$\gamma$ experiment \cite{GEp2gamma} was aimed at observing $\ve$ dependence of proton FF ratio\footnote{%
Throughout the paper we denote $R = G_E/G_M$, but in the literature another definition is frequently used: $R = \mu G_E/G_M$, such that
$\left. R \right|_{Q^2\to 0} = 1$.}
$R = G_E/G_M$, measured via polarization transfer method. Such a dependence should constitute clear sign of TPE. In terms of invariant amplitudes $\CG$, the TPE correction to this ratio is
\be
 \frac{\delta R_\text{exp}}{R_\text{exp}} = \Re \left\{ \frac{\delta\CG_E}{G_E}
    - \frac{\delta\CG_M}{G_M}
    - \frac{\ve(1-\ve)}{1+\ve} \frac{\delta\CG_3}{G_M} \right\}
\ee
The data were taken at fixed $Q^2 = 2.5 \GeV^2$ and three different values of $\ve$.
Also $S_\para$, the longitudinal polarization component of the final proton%
\footnote{Denoted $P_l$ in original papers.},
was measured, where TPE correction is equal [cf. Eq.~(\ref{PT2g})]:
\be
 \frac{\delta S_\para}{S_\para} = - 2\ve \Re \left\{
   \frac{R^2}{\ve R^2 + \tau}
     \left( \frac{\delta\CG_E}{G_E} - \frac{\delta\CG_M}{G_M} \right)
   + \frac{\ve}{1+\ve} \frac{\delta\CG_3}{G_M}
 \right\}
\ee

\begin{figure}
\centering
\includegraphics[width=0.45\textwidth]{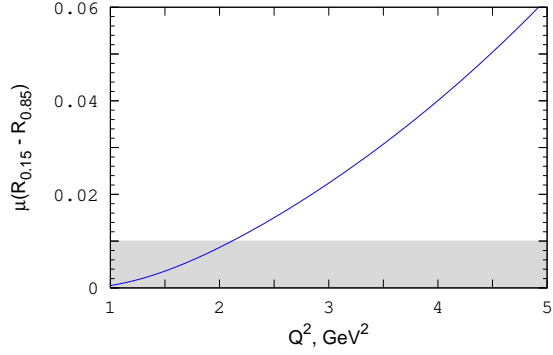}
\caption{A difference between TPE correction to measured $\mu R = \mu G_E/G_M$ at $\ve=0.15$ and $\ve=0.85$, versus $Q^2$, calculated according to \cite{ourPiN}. Gray band at the bottom shows typical error of GEp-2$\gamma$ experiment.}\label{deltaRrange}
\end{figure}

Somewhat surprisingly, no $\ve$ dependence of $R$ was seen in the experiment: the three measured values are almost the same (within 1.5\% bounds).
One possible cause for such outcome is that near this kinematical point the variation of total (elastic + inelastic) TPE correction to $G_E/G_M$ is minimal (Fig.~\ref{deltaRrange}). An experiment at higher $Q^2$ value (for instance $Q^2 = 3.5\GeV^2$) with the same accuracy would clearly reveal $\ve$ dependence caused by TPE.

For $S_\para$, some rise (about 2.3\%) at higher $\ve$ was observed, but it should be noted that the point with the smallest $\ve$ was used for normalization here, and essentially we remain with only two points, so it is hard to make any conclusions about the trend.

Several experiments were dedicated to measuring charge asymmetry, i.e. $e^+p/e^-p$ scattering cross-section ratio. Since the TPE correction has opposite signs for $e^+p$ and $e^-p$, the deviation of the ratio from unity is a direct TPE effect, and equal twice the cross-section correction:
\be
 1 - R_\pm = 2\frac{\delta\sigma}{\sigma} = 
 \frac{4}{\ve R^2 + \tau} \Re \left\{ \ve R^2 \frac{\delta\CG_E}{G_E} + \tau \frac{\delta\CG_M}{G_M} \right\}
\ee

In VEPP-3 experiment \cite{VEPP}, the ratio $R_\pm$ was measured at beam energy $E=1.0$ and $1.6 \GeV$, and several different angles, which correspond to $Q^2$ from 0.8 to $1.5 \GeV^2$ and $\ve$ from 0.27 to 0.45. The measured ratio $R_\pm$ was clearly larger than unity and agree well with theoretical calculations \cite{ourPi}.

In CLAS experiment \cite{CLAS}, two series of measurements were performed: at fixed $Q^2 = 1.45 \GeV^2$ and $\ve$ from 0.4 to 0.9 and at fixed $\ve = 0.88$ and $Q^2$ from 0.2 to $1.4 \GeV^2$. Though the results are rather close to 1 within errors, they show $2.5\sigma$ preference to TPE over ``no TPE'' \cite{CLAS}; in Ref.~\cite{ourPiN} it was shown that full TPE contribution (elastic+inelastic) is best consistent with data (has least $\chi^2$), the elastic only is slightly worse and the worst is absence of TPE.

In the OLYMPUS experiment \cite{OLYMPUS}, the ratio was measured with $E=2 \GeV$ electrons and positrons at $Q^2$ from 0.6 to $2.2 \GeV^2$ ($\ve$ from 0.4 to 0.9).
It is interesting that preliminary results in the region $\ve \approx 0.8-0.9$ are below unity and thus contradict theoretical predictions from Refs.~\cite{bmt,ourDisp,ourPiN}.

%%%==================================================================================
\section{Extraction of TPE amplitudes}\label{sec:extraction}
When checking for TPE effects, another approach is possible: assuming that the scattering amplitude is a sum of OPE and TPE,
and the discrepancy between the Rosenbluth and polarization methods is entirely due to TPE,
one may try to extract the TPE amplitude directly from the experimental data.
Thus extracted value will be model-independent and can be easily compared to the theoretical calculations.

The first attempts to do this were carried out in Refs.\cite{GV,ArringtonExtr}.
However, authors of that works arbitrarily assumed that only one of three TPE amplitudes, $Y_{2\gamma}$ ($\delta\CG_3/G_M$ in our notation)
is responsible for the discrepancy.
%% почему и результаты получились неправильные

The rigorous application of such an approach was done in \cite{ourPheno}. Using the generalized FFs, introduced in Eq.(\ref{amplCG}), one can see that at high and even at moderate $Q^2$ the TPE correction to the cross-section comes almost entirely from the magnetic FF:
\be
 \sigma_R = \ve |\CG_E|^2 + \tau |\CG_M|^2 = 
	\ve G_E^2 + 2 \ve G_E \Re \delta\CG_E + \tau \left[ G_M^2 + 2 G_M \Re \delta\CG_M \right]
\ee
but the term with $\delta\CG_E$ is the smallest one, since not only $\tau \equiv \frac{Q^2}{4M^2} \gg \ve$, but also $G_E/G_M \approx 1/\mu_p \approx 1/2.79$, thus
\be
 \sigma_R \approx \ve G_E^2(Q^2) + \frac{Q^2}{4M^2} \left[ G_M^2(Q^2) + 2 G_M(Q^2) \Re \delta\CG_M(Q^2, \ve) \right]
\ee
Given the experimental fact that the Rosenbluth plots are linear (i.e. reduced cross-section is linear function of $\ve$),
one may conclude that $\delta\CG_M$ should be approximately linear function of $\ve$.
Additionally recalling that the the dispersion relations require $\left. \delta\CG_M \right|_{\ve=1} = 0$, we arrive at
\be
 \delta\CG_M(Q^2, \ve) = a(Q^2) (1 - \ve)
\ee
As a result, $R_{LT}$, the FF ratio, measured with Rosenbluth method, become equal
\be
 R_{LT}^2 = \frac{R^2 - 2a\tau}{\tau(1+2a)}
\ee
where $R = G_E/G_M$ is true FF ratio, and we can determine the coefficient $a$ and the whole $\delta\CG_M$ amplitude directly from the combination of Rosenbluth and polarization transfer data
(the last point was missed in Refs.~\cite{ourPheno}, but corrected in Ref.~\cite{ourPheno2}).

\begin{figure}
\centering
\includegraphics[width=0.45\textwidth]{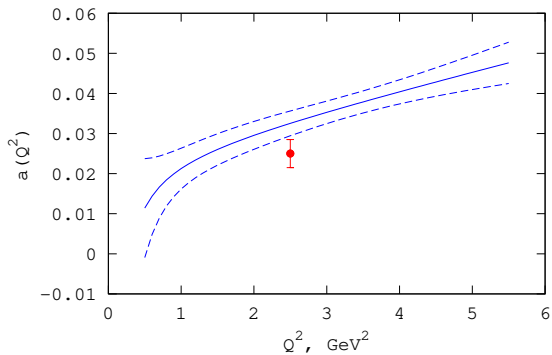}
\caption{Coefficient $a$, as determined in Refs.~\cite{ourPheno} (blue lines, $\pm 1\sigma$) and \cite{ourPheno2} (red point).}\label{phenoPlot}
\end{figure}

After the GEp-2$\gamma$ experiment \cite{GEp2gamma}, new precise data at $Q^2 = 2.5 \GeV^2$ became available, and a new analysis was carried out in Refs. \cite{ourPheno2} and \cite{KivelPheno}.

In Ref.~\cite{ourPheno2} the slope of $\delta\CG_M$ was determined from new data and again turned out to be in agreement with theoretical calculations.
Another quantity, measured in the experiment (longitudinal component of the final proton polarization, $S_\para$), gave an opportunity to determine $\delta\CG_3$, though the precision was much worse here. In fact, the experimental value is compatible with zero.
The third amplitude $\delta\CG_E$ could not be determined accurately, but basing on the fact that experimental $R$ varies very little with $\ve$ one can conclude that $\delta\CG_E/G_E \approx \delta\CG_M/G_M$.

Similar program was tried by the authors of Ref.~\cite{KivelPheno}; to extract TPE amplitudes they assumed a custom $\ve$ dependence of the final proton polarization $S_\para$. Nevertheless, their conclusion was that only one of three TPE amplitudes could be extracted reliably.

%%%==================================================================================
\section{TPE in other processes}\label{sec:other}

The previous sections were devoted mainly to elastic electron-proton scattering
(or, possibly, electron-nucleon scattering, since the neutron differs from the proton only in FFs).
In this section, we will consider TPE in other processes.
First, electron can be replaced with muon to see effects, related to lepton mass,
and then we consider electron scattering off different targets, in particular pions and light nuclei (deuteron, $^3$He, $^3$H).
Other processes, for which TPE effects were studied, but will not be discussed here, include electroproduction of pions~\cite{PionProd} and $\Delta$ resonance~\cite{DeltaProd}, parity-violating $ep$ scattering~\cite{TBE} and deep inelastic scattering~\cite{TPE-DIS}.

\subsection{Muon-proton scattering}\label{sec:muon}

After discovery of the proton radius puzzle, one of the possible explanations
was that the e.m. interaction of the muon is in some way different from that of the electron.
To test this possibility, it was proposed to measure proton e.m. radius
in the scattering approach, but with the muon beam \cite{MUSE}.

In order to handle the results of such experiment, one should be able to calculate
TPE corrections to $\mu p$ scattering. They were studied in Refs.~\cite{Tomalak-mup} and \cite{AfanasevTri,ourTri}.
The main difference between $ep$ and $\mu p$ scattering is that the muon mass is not negligible.
Therefore, there are six, not three, invariant amplitudes now:
\be
  \CM = - \frac{4\pi\alpha}{q^2} \left\{
     \, \bar u'\gamma_\mu u \, \bar U' \gamma_\nu U
  \left[   (F_1+F_2) g_{\mu\nu}
          - F_2 \frac{P_\mu P_\nu}{M^2}
          + F_3 \frac{P_\mu K_\nu}{M^2}
          + F_4 \frac{K_\mu P_\nu}{m^2}
          - F_5 \frac{K_\mu K_\nu}{m^2}
  \right]
          - F_6 \, \bar u'\gamma_5 u \, \bar U' \gamma_5 U
   \right\},    
\ee
The properties of the parameter $\ve$ in the Rosenbluth formula also change: instead of (\ref{epsilon}) it now has the form
\be
\ve = \frac{\nu^2 - Q^2(4M^2+Q^2)}{\nu^2 + (Q^2-4m^2)(4M^2+Q^2)}
\ee
and, for fixed $Q^2$, varies not from 0 to 1, but between $2m^2/Q^2$ and 1; near-forward scattering, as before, corresponds to $\ve \to 1$.

The elastic contribution to the TPE in $\mu p$ scattering for low momentum transfer was calculated in Ref.~\cite{Tomalak-mup}.
The same method as described in Sec.\ref{sec:elastic} was used.
It was found that TPE correction to the cross-section is about 0.5\% and several times smaller than for $ep$ scattering with similar kinematics,
because the contribution of the lepton-spin-flipping amplitudes ($F_4$ and $F_5$; the amplitude $F_6$ does not contribute to the unpolarized cross-section) has different sign and partially cancel spin-nonflipping amplitudes.

Later, the same authors developed a dispersion formalism to calculate TPE amplitudes for $\mu p$ scattering \cite{Tomalak-disp-subtracted, Tomalak-disp-muon}. It is interesting to note that the calculation of amplitude $F_4$ requires a subtracted dispersion relation.

\subsubsection{Triangle diagram}
\begin{figure}
 \centering
 \includegraphics[width=0.16\textwidth]{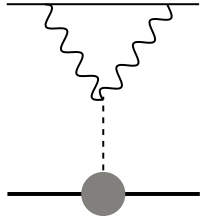}
 \caption{Triangle diagram.} \label{diag-tri}
\end{figure}
In Refs.~\cite{AfanasevTri,ourTri}, completely new type of TPE contribution (first mentioned in Ref.~\cite{ZhouTri}) was considered: the triangle diagram, Fig.~\ref{diag-tri}.
It arises from the exchange of a single $C$-even meson, which then decays in two virtual photons. Certainly, it exists for the $ep$ scattering as well, but is negligible there, because its contribution is proportional to the lepton mass. On the other hand, for the $\mu p$ scattering it cannot be left out, since the muon mass is three orders of magnitude larger.

It it natural to expect that larger contributions would come from the lightest mesons, as the expression for the diagram contains
the factor $1/(Q^2+\mu^2)$, originating from the meson propagator (where $\mu$ is meson mass).
The lightest candidate is $\pi$ meson, but it was shown in Ref.~\cite{AfanasevTri} that the contribution of a pseudoscalar meson does not interfere with the OPE amplitude and thus has no effect in unpolarized scattering.
The next lightest meson is scalar $\sigma(500)$. Its contribution was estimated to be about 0.1\% in kinematics of MUSE experiment \cite{MUSE}.
It was also found that the contribution of the triangle diagram strongly grows at $Q^2\to 0$.

In Ref.~\cite{ourTri} low-$Q^2$ behaviour of the triangle diagram and its the possible effect of proton radius puzzle was studied. It was shown that the corresponding amplitude follows approximately $a + b\sqrt{Q^2}$ formula, and the shift of the muon hydrogen energy levels, induced by the diagram, is about 30 times smaller than needed to resolve the discrepancy between different measurements of the proton radius.

\subsection{TPE on nuclei}

The TPE in the elastic electron scattering off the light nuclei ($d$, $^3$He, $^3$H) was studied in Refs.~\cite{Chin-deut,Chin-deut-2,Kob-deut,Kob-deut-2,Kob-tri}.

The important point here is that the nuclear elastic FFs decline with $Q^2$ rather rapidly (exponentially)
and therefore the significance of TPE may be enhanced.

If we consider elastic contribution only, then the nucleus is viewed as a single particle.
However, nuclear excited states are usually much closer to the ground state than for the proton,
thus the inelastic contributions are probably important.

If we want go further and consider inelastic contributions as well, we must take into account the internal structure of the nucleus.
At moderate energies it is quite natural to view the nucleus as composed of nucleons and neglect quark degrees of freedom.
Thus, similarly to pQCD picture for proton, there are two types of Feynman diagrams here: where both virtual photons interact with the same nucleon (type I) and where photons interact with different nucleons (type II). In such calculations, the motion of the nucleons inside the nucleus is usually described in the nonrelativistic or semirelativistic approximation.

\subsubsection{Spin-1/2 ($^3$He and $^3$H)}
The $^3$He and $^3$H nuclei consist of three nucleons and have spin 1/2, similarly to the proton.
Thus the general formula for the TPE amplitude (\ref{genAmp}) holds here as well: the are three independent TPE amplitudes.

The elastic contribution (i.e. ignoring excitation of the nucleus) to the TPE for electron scattering off $^3$He target was calculated in Refs.~\cite{bmt} and \cite{ourNR} using the same methods which were developed for the electron-proton scattering.

More thoroughly TPE amplitudes for the elastic electron-trinucleon scattering were studied in Ref.~\cite{Kob-tri} %
using semirelativistic nuclear wavefunction, corresponding to popular Paris and CD-Bonn nucleon-nucleon interaction potentials. It was found that the TPE corrections are several times larger than in electron-proton scattering, and the diagrams of type II are dominant, except the correction to the magnetic FF at large $Q^2$.

\subsubsection{Deuteron}
The deuteron has spin 1 and thus the structure of the scattering amplitude is different from the previous (spin-1/2) case.
For the electron-deuteron scattering, in OPE approximation there are three FFs: electric $G_C$, quadrupole $G_Q$ and magnetic $G_M$; and in general case there are six.

It was shown \cite{Kob-deut} that TPE can give large contributions (up to 10-20\% at $Q^2 > 2.5\GeV^2$ \cite{Kob-deut-2}) to the elastic $ed$ scattering cross-section, but it largely depends on the deuteron wave function at short distances, which is poorly known.
It was also shown that tensor analyzing power component $T_{22}$ is mainly determined by TPE at $Q^2 > 0.5\GeV^2$ \cite{Kob-deut-2}.

TPE in elastic $ed$ scattering was also studied in the framework of effective Lagrangian approach \cite{Chin-deut, Chin-deut-2}. The results were somewhat different, in particular, it was found that the largest effect of TPE is seen in the polarization observable $T_{10}$ at small scattering angles.

\subsection{TPE on pion}
TPE in elastic electron-pion scattering is significantly simpler to study than in the electron-proton case.
Since the pion is spin-0 particle, for both OPE and TPE there is only one invariant amplitude, or FF, $F(q^2)$:
\be
  \CM = -\frac{4\pi\alpha}{q^2} \bar u'\gamma_\mu u \, (p+p')_\mu \, F(q^2)
\ee
Thus the only effect of TPE is a correction to this FF (and its dependence on $\ve$).
The first calculations of this correction was done in Refs.\cite{bmtPi,otherPi}.
In both papers only the elastic contribution was considered;
authors of Ref.~\cite{otherPi} additionally assume that each of the virtual photos carries about a half of the transferred momentum,
while authors of Ref.~\cite{bmtPi} perform the full calculation of box and crossed-box diagrams via n-point functions (the approach, similar to Ref.~\cite{bmt}).
Numerically, the correction was found to be about 1\%, smaller at small $Q^2$ and increasing at $Q^2 \gtrsim 1\GeV^2$, especially sharply  at extreme backward angles.

It was argued in Ref.~\cite{ZhouPi}, that the virtual Compton scattering tensor, implicitly used in these works,
breaks gauge invariance, and the so-called contact term should be added to it.
Numerically, however, the change turned out to be small.

In Ref.~\cite{ourPi} TPE correction to the pion FF was calculated using appropriately modified dispersion approach of Ref.~\cite{ourDisp} with elastic and inelastic ($\rho$ and $b_1$ mesons) intermediate states.
The results for the elastic contribution were in agreement with previous works.
The inelastic contribution was found to be negligible with respect to the elastic one for small $Q^2$, similarly to the $ep$ case,
and becoming comparable to it at $Q^2 \gtrsim 2\GeV^2$, though being still smaller.
It was also shown that the sharp growth of the high-$Q^2$ TPE amplitude at backward angles is due to the $u$-channel threshold singularity, which at $Q^2 \gg M^2$ is close to the boundary of the physical region.

%%%==================================================================================
\section{Summary and outlook}
Since the discrepancy in proton FF measurements was revealed in mid-2000s,
our understanding of TPE and its role in elastic $ep$ scattering has advanced significantly.
The methods were proposed to calculate TPE amplitudes at low and intermediate $Q^2$,
including elastic and simplest inelastic intermediate hadronic states. The discrepancy is most likely explained by TPE; TPE amplitudes, extracted from the combination of Rosenbluth and polarization transfer data, agree with theoretical calculations.

At low $Q^2$ the TPE corrections influence proton radius measurements,
both by $ep$ scattering and by Lamb shift in muonic hydrogen, and should be taken into account in such experiments.
For high-$Q^2$ kinematics, QCD-based methods of TPE calculations are available.

Several dedicated experiments on $e^+/e^-$ charge asymmetry in elastic scattering show clear preference to ``TPE'' over ``no TPE'' hypothesis, but further, more precise, measurements would be in order.

The failure to observe the expected TPE effect in GEp-2$\gamma$ experiment is likely explained
by the unfortunate choice of kinematics.
We suggest conducting similar experiment at higher $Q^2$ (for instance, $3.5 \GeV^2$),
which should be sufficient to see clear $\ve$ dependence of polarization ratio.

In recent years, theoretical understanding of TPE in elastic muon-proton and electron-nucleus scattering was actively developing and some work is still to be done in these fields.

\appendix
\section* {Appendix: TPE corrections to observables}
In this section, the formulae for the TPE contributions to various observables of elastic electron-proton scattering are gathered.
The electron mass is neglected.
The amplitudes $\delta\CG_E$, $\delta\CG_M$ and $\delta\CG_3$ are defined according to Eqs.~(\ref{amplCG}, \ref{genAmp}). In the following equations, $R = G_E/G_M$ and $\tau = Q^2/4M^2$.

The unpolarized cross-section and electron-positron cross-section ratio:
\be
 \frac{\delta\sigma}{\sigma} = \frac{1- R_\pm}{2} = 
 \frac{2}{\ve R^2 + \tau} \Re \left\{ \ve R^2 \frac{\delta\CG_E}{G_E} + \tau \frac{\delta\CG_M}{G_M} \right\}
\ee

The FF ratio $R_\text{exp}$, as measured in polarization transfer method:
\be
 \frac{\delta R_\text{exp}}{R_\text{exp}} = \Re \left\{ \frac{\delta\CG_E}{G_E}
    - \frac{\delta\CG_M}{G_M}
    - \frac{\ve(1-\ve)}{1+\ve} \frac{\delta\CG_3}{G_M} \right\}
\ee

The longitudinal component of final proton polarization:
\be
 \frac{\delta S_\para}{S_\para} = - 2\ve \Re \left\{
   \frac{R^2}{\ve R^2 + \tau}
     \left( \frac{\delta\CG_E}{G_E} - \frac{\delta\CG_M}{G_M} \right)
   + \frac{\ve}{1+\ve} \frac{\delta\CG_3}{G_M}
 \right\}
\ee

Target normal spin asymmetry:
\be
 A_n = - \sqrt{2\ve(1+\ve)} \frac{R\sqrt{\tau}}{\ve R^2 + \tau}
 \Im \left\{ \frac{\delta\CG_E}{G_E} - \frac{\delta\CG_M}{G_M}
     - \frac{\ve(1-\ve)}{1+\ve} \, \frac{\delta\CG_3}{G_M} \right\}.
\ee


\begin{thebibliography}{150}
\bibitem{Hof1}
  \Jou{R.~Hofstadter, R.W.~McAllister}{Phys. Rev.}{98}{217}{1955}{}
\bibitem{Hof2}
  \Jou{E.~Chambers, R.~Hofstadter}{Phys. Rev.}{103}{1454}{1956}{} 
\bibitem{Hof3}
  \Jou{R.~Hofstadter}{Rev. Mod. Phys.}{28}{214}{1956}{}
\bibitem{CODATA}
  \Jou{R.J.~Mohr, D.B.~Newell, B.N.~Taylor}{Rev. Mod. Phys}{88}{035009}{2016}{} 
\bibitem{rE-muon-0}
  \Jou{R.~Pohl \ea}{Nature}{466}{213}{2010}{}
\bibitem{rE-muon}
  \Jou{A. Antognini \ea}{Science}{339}{417}{2013}{}
\bibitem{PRad}
  \Jou{W. Xiong \ea}{Nature}{575}{147-150}{2019}
  {A small proton charge radius from an electron–proton scattering experiment}

\bibitem{Hof4}
  \Jou{M.R.~Yearian, R.~Hofstadter}{Phys. Rev.}{110}{553}{1958}{}

\bibitem{Rosenbluth}
  \Jou{M.N.~Rosenbluth}{Phys. Rev.}{79}{615-619}{1950}
  {High Energy Elastic Scattering of Electrons on Protons}
\bibitem{Jones}
  \Jou{M.K.~Jones \ea}{Phys. Rev. Lett.}{84}{1398}{2000}{}
\bibitem{Punjabi}
  \Jou{V.~Punjabi \ea}{Phys. Rev. C}{71}{055202}{2005}{}
\bibitem{Gayou}
  \Jou{O.~Gayou \ea}{Phys. Rev. Lett.}{88}{09230}{2002}{}
\bibitem{Puckett}
  \Jou{A.~Puckett \ea}{Phys. Rev. C}{85}{045203}{2012}{}
\bibitem{bmt0}
  \Jou{P.G.~Blunden, W.~Melnitchouk, J.A.~Tjon}{Phys. Rev. Lett.}{91}{142304}{2003}
  {Two-Photon Exchange and Elastic Electron-Proton Scattering}
\bibitem{GV}
  \Jou{P.A.M.~Guichon, M.~Vanderhaeghen}{Phys. Rev. Lett.}{91}{142303}{2003}
  {How to Reconcile the Rosenbluth and the Polarization Transfer Methods in the Measurement of the Proton Form Factors}
\bibitem{Sachs}
  \Jou{R.G.~Sachs}{Phys. Rev.}{126}{2256-2260}{1962}
  {High-Energy Behavior of Nucleon Electromagnetic Form Factors}
\bibitem{ARekalo1}
  \Jou{А.И.~Ахиезер, М.П.~Рекало}{Докл. АН СССР}{180, \No 5}{1081-1083}{1968}
  {Поляризационные явления при рассеянии электронов протонами в области больших энергий}
\bibitem{ARekalo2}
  \Jou{А.И.~Ахиезер, М.П.~Рекало}{ЭЧАЯ}{4, \No 3}{662-688}{1973}
  {Поляризационные явления при рассеянии лептонов адронами}
\bibitem{Milbrath}
  \Jou{B.D.~Milbrath \ea}{Phys. Rev. Lett.}{80}{452-455}{1998}
  {Comparison of Polarization Observables in Electron Scattering from the Proton and Deuteron}
  Erratum --- \Jou{B.D.~Milbrath \ea}{Phys. Rev. Lett.}{82}{2221}{1999}{}
\bibitem{PTJones}
  \Jou{M.K.~Jones \ea}{Phys. Rev. C}{74}{035201}{2006}
  {Proton $G_{E}/G_{M}$ from beam-target asymmetry}
\bibitem{LTWalker}
  \Jou{R.C.~Walker \ea}{Phys. Rev. D}{49}{5671-5689}{1994}
  {Measurements of the proton elastic form factors for $1 \le Q^2 \le 3\ (\GeV/c)^2$ at SLAC}
\bibitem{LTAndivahis}
  \Jou{L.~Andivahis \ea}{Phys. Rev. D}{50}{5491-5517}{1994}
  {Measurements of the electric and magnetic form factors of the proton from $Q^2=1.75$ to $8.83\ (\GeV/c)^2$}
\bibitem{LTChristy}
  \Jou{M.E.~Christy \ea}{Phys. Rev. C}{70}{015206-1-15}{2004}
  {Measurements of electron-proton elastic cross sections for  $0.4 < Q^2 < 5.5\ (\GeV/c)^2$}

\bibitem{LTQattan}
  \Jou{I.A.~Qattan \ea}{Phys. Rev. Lett.}{94}{142301}{2005}
  {Precision Rosenbluth Measurement of the Proton Elastic Form Factors}
\bibitem{Puckett2017}
  \Jou{A.~Puckett \ea}{Phys. Rev. C}{96}{055203}{2017}{}

\bibitem{GEp2gamma}
  \Jou{M. Meziane \ea}{Phys. Rev. Lett.}{106}{132501}{2011}{}
\bibitem{ArringtonFit}
  \Jou{J.~Arrington,W.~Melnitchouk,J.A.~Tjon}{Phys. Rev. C}{76}{035205}{2007}
  {Global analysis of proton elastic form factor data with two-photon exchange corrections}
\bibitem{Arrington}
  \Jou{J.~Arrington}{Phys. Rev. C}{68}{034325}{2003}
  {How well do we know the electromagnetic form factors of the proton?}
\bibitem{QCD1}
  \Jou{G.P.~Lepage, S.J.~Brodsky}{Phys. Rev. D}{23}{2157-2198}{1980}
  {Exclusive processes in perturbative quantum chromodynamics}
\bibitem{QCD2}
  \Jou{В.Л.~Черняк, А.Р.~Житницкий}{ЯФ}{31, \No 4}{1053-1068}{1980}
  {Асимптотика адронных формфакторов в квантовой хромодинамике}
\bibitem{Landau}
  {\it Берестецкий~В.Б., Лифшиц~Е.М., Питаевский~Л.П.}
  Релятивистская квантовая теория, ч.1. -- М.: Наука, 1968. -- 480~с.
\bibitem{MoTsai}
  \Jou{L.W.~Mo, Y.S.~Tsai}{Rev. Mod. Phys.}{41}{205-235}{1969}
  {Radiative Corrections to Elastic and Inelastic ep and up Scattering}
\bibitem{Tsai}
  \Jou{Y.S.~Tsai}{Phys. Rev.}{122}{1898-1907}{1961}
  {Radiative Corrections to Electron-Proton Scattering}
\bibitem{AfaBS1}
  \Jou{A. Afanasev, I. Akushevich, N. Merenkov}{Phys. Rev. D}{64}{113009}{2001}
  {Model independent radiative corrections in processes of polarized electron nucleon elastic scattering}
\bibitem{AfaBS2}
  \Jou{A. Afanasev, I. Akushevich, A. Ilyichev, N. Merenkov}{Phys. Lett. B}{514}{269-278}{2001}
  {QED radiative corrections to asymmetries of elastic ep scattering in hadronic variables}
\bibitem{ourBS}
  \Jou{D. Borisyuk, A. Kobushkin}{Phys. Rev. C}{90}{025209}{2014}
  {Radiative corrections to polarization observables in electron-proton scattering}

\bibitem{DrellFu}
  \Jou{S.D.~Drell, S.~Fubini}{Phys. Rev.}{113}{741-744}{1959}
  {Higher Electromagnetic Corrections to Electron-Proton Scattering}
\bibitem{Greenhut}
  \Jou{G.K.~Greenhut}{Phys. Rev.}{184}{1860-1867}{1969}
  {Two-photon exchange in electron-proton scattering}

\bibitem{deRuj1}
  \Jou{A.~De Rujula, J.M.~Kaplan, E.~de Rafael}{Nucl. Phys.}{B35}{365-389}{1971}
  {Elastic scattering of electrons from polarized protons and inelastic electron scattering experiments}
\bibitem{deRuj2}
  \Jou{A. De Rujula, J.M. Kaplan, E. De Rafael}{Nucl. Phys.}{B53}{545-566}{1973}
  {Optimal positivity bounds to the up-down asymmetry in elastic electron-proton scattering}
\bibitem{PV}
  \Jou{B.~Pasquini, M.~Vanderhaeghen}{Phys. Rev. C}{70}{045206}{2004}
  {Resonance estimates for single spin asymmetries in elastic electron-nucleon scattering}
\bibitem{ourTNSA}
  \Jou{D.~Borisyuk, A.~Kobushkin}{Phys. Rev. C}{72}{035207}{2005}
  {Target normal spin asymmetry of the elastic ep scattering at resonance energy}
\bibitem{MAID}
 \Jou{D. Drechsel, S.S. Kamalov, L. Tiator}{Eur. Phys. J. A}{34}{69-97}{2007}{}
\bibitem{AfanMerCorr}
  \Jou{A.~Afanasev, N.P.~Merenkov}{Phys.Lett.B}{599}{48}{2004}
  {Collinear Photon Exchange in the Beam Normal Polarization Asymmetry of Elastic Electron-Proton Scattering}

\bibitem{Gorchtein}
  \Jou{M.~Gorchtein}{Phys. Rev. C}{73}{055201}{2006}
  {Beam normal spin asymmetry in the quasi-RCS approximation}

\bibitem{ourBNSA}
  \Jou{D.~Borisyuk, A.~Kobushkin}{Phys. Rev. C}{73}{045210}{2006}
  {Beam normal spin asymmetry of elastic eN scattering in the leading logarithm approximation}
\bibitem{mue-mue}
  \Jou{P.~van Nieuwenhuizen}{Nucl. Phys.}{B28}{429-454}{1971}
  {Muon-electron scattering cross section to order $\alpha^3$}
\bibitem{bmt}
  \Jou{P.G.~Blunden, W.~Melnitchouk, J.A.~Tjon}{Phys. Rev. C}{72}{034612}{2005}
  {Two-Photon Exchange in Elastic Electron-Nucleon Scattering}
\bibitem{tHooft}
  \Jou{G. 't Hooft, M.J.G. Veltman}{Nucl. Phys.}{B153}{365-401}{1979}
  {Scalar One Loop Integrals}

\bibitem{MaximonTjon}
  \Jou{L.C. Maximon, J.A. Tjon}{Phys. Rev. C}{62}{054320}{2000}
  {Radiative corrections to electron proton scattering}
\bibitem{ourBox}
  \Jou{D.~Borisyuk, A.~Kobushkin}{Phys. Rev. C}{74}{065203}{2006}
  {Box diagram in the elastic electron-proton scattering}
\bibitem{ourDisp}
  \Jou{D. Borisyuk, A. Kobushkin}{Phys. Rev. C}{78}{025208}{2008}{}
\bibitem{bmtDelta}
  \Jou{S.~Kondratyuk, P.G.~Blunden, W.~Melnitchouk, J.A.~Tjon}{Phys. Rev. Lett.}{95}{172503}{2005}
  {$\Delta$ Resonance Contribution to Two-Photon Exchange in Electron-Proton Scattering}

\bibitem{bmtRes}
  \Jou{S. Kondratyuk, P.G. Blunden}{Phys. Rev. C}{75}{038201}{2007}{}
\bibitem{ourBoxRes}
  {\it D.~Borisyuk, A.~Kobushkin}, unpublished.
\bibitem{ourDelta}
  \Jou{D. Borisyuk, A. Kobushkin}{Phys. Rev. C}{86}{055204}{2012}{}
\bibitem{ourP33}
  \Jou{D. Borisyuk, A. Kobushkin}{Phys. Rev. C}{89}{025204}{2014}{}
\bibitem{ourPiN}
  \Jou{D. Borisyuk, A. Kobushkin}{Phys. Rev. C}{92}{035204}{2015}{}

\bibitem{GPD-rev}
  \Jou{M. Tanabashi \ea{} (Particle Data Group)}{Phys. Rev. D}{98}{030001}{2018}
  {18. Structure Functions, 18.6. Generalized parton distributions}
\bibitem{GPD}
  \Jou{Y.C.~Chen, \ea}{Phys. Rev. Lett.}{93}{122301}{2004}
  {Partonic Calculation of the Two-Photon Exchange Contribution to Elastic Electron-Proton Scattering at Large Momentum Transfer}
  \Jou{A.V.~Afanasev \ea}{Phys. Rev. D}{72}{013008}{2005}
  {The Two-photon exchange contribution to elastic electron-nucleon scattering at large momentum transfer}
\bibitem{ourQCD}
  \Jou{D.~Borisyuk, A.~Kobushkin}{Phys. Rev. D}{79}{034001}{2009}{}
\bibitem{KivelQCD}
  \Jou{N. Kivel, M. Vanderhaeghen}{Phys. Rev. Lett.}{103}{092004}{2009}
  {Two-photon exchange in elastic electron-proton scattering: QCD factorization approach}

\bibitem{SCET}
  \Jou{N.~Kivel, M.~Vanderhaeghen}{JHEP}{1304}{029}{2013}
  {Two-photon exchange corrections to elastic electron-proton scattering at large momentum transfer within the SCET approach}

\bibitem{Sick}
  \Jou{I.~Sick}{Phys. Lett.}{B576}{62-67}{2003}
  {On the RMS radius of the proton}
\bibitem{McKinley}
  \Jou{W.A.~McKinley, H.~Feshbach}{Phys. Rev.}{74}{1759-1763}{1948}
  {The Coulomb Scattering of Relativistic Electrons by Nuclei}
\bibitem{Dalitz}
  \Jou{R.H.~Dalitz}{Proc. Roy. Soc. Lond.}{A206}{509-520}{1951}
  {On higher Born approximations in potential scattering}
\bibitem{Rosenfelder}
  \Jou{R.~Rosenfelder}{Phys. Lett.}{B479}{381-386}{2000}
  {Coulomb corrections to elastic electron-proton scattering and the proton charge radius}
\bibitem{Lewis}
  \Jou{R.R. Lewis, Jr.}{Phys. Rev.}{102}{537}{1956}{}
\bibitem{ourLow}
  \Jou{D.~Borisyuk, A.~Kobushkin}{Phys. Rev. C}{75}{038202}{2007}
  {Two-photon exchange at low $Q^2$}
\bibitem{ourNR}
  {\it D. Borisyuk, A. Kobushkin}, arXiv:1811.06928.

\bibitem{BlundenSick}
  \Jou{P.G.~Blunden, I.~Sick}{Phys. Rev. C}{72}{057601}{2005}
  {Proton radii and two-photon exchange}
\bibitem{ourRadius}
  \Jou{D.~Borisyuk}{Nucl. Phys.}{A843}{59-67}{2010}
  {Proton charge and magnetic rms radii from the elastic ep scattering data}
\bibitem{rE-electron}
  \Jou{P.J. Mohr, B.N. Taylor, D.B. Newell}{Rev. Mod. Phys.}{84}{1527}{2012}{}
\bibitem{MUSE}
  {\it R. Gilman \ea}, arXiv:1303.2160.

\bibitem{Bernauer}
  \Jou{J.C.~Bernauer \ea}{Phys. Rev. Lett.}{105}{242001}{2010}{}
\bibitem{Bernauer2}
  \Jou{J.C.~Bernauer \ea}{Phys. Rev. C}{90}{015206}{2014}
  {Electric and magnetic form factors of the proton}
\bibitem{VEPP}
  \Jou{I.A. Rachek \ea}{Phys. Rev. Lett.}{114}{062005}{2015}{}
\bibitem{ourPi}
  \Jou{D.~Borisyuk, A.~Kobushkin}{Phys. Rev. C}{83}{025203}{2011}{}

\bibitem{CLAS}
  \Jou{D. Adikaram \ea}{Phys. Rev. Lett.}{114}{062003}{2015}{}
\bibitem{OLYMPUS}
  \Jou{Brian S. Henderson}{PoS}{310}{149}{2018}
  {Results from the OLYMPUS Experiment on the Contribution of Hard Two-Photon Exchange to Elastic Electron-Proton Scattering} 
\bibitem{ArringtonExtr}
  \Jou{J. Arrington}{Phys. Rev. C}{71}{015202}{2005} 
  {Extraction of two-photon contributions to the proton form-factors}
\bibitem{ourPheno}
  \Jou{D.~Borisyuk, A.~Kobushkin}{Phys. Rev. C}{76}{022201}{2007}
  {Phenomenological analysis of two-photon exchange effects in proton form factor measurements}
\bibitem{ourPheno2}
  \Jou{D.~Borisyuk, A.~Kobushkin}{Phys. Rev. D}{83}{057501}{2011}
  {Two-photon exchange amplitudes for the elastic ep scattering at Q^2=2.5 GeV^2 from the experimental data}
\bibitem{KivelPheno}
  \Jou{J.~Guttmann, N.~Kivel, M.~Meziane, M.~Vanderhaeghen}{Eur. Phys. J. A}{47}{77}{2011}
  {Determination of two-photon exchange amplitudes from elastic electron-proton scattering data}

\bibitem{PionProd}
  \Jou{A. Afanasev, A. Aleksejevs, S. Barkanova}{Phys.Rev. D}{88}{053008}{2013}
   {Two Photon Exchange for Exclusive Pion Electroproduction}
\bibitem{DeltaProd}
  \Jou{S.~Kondratyuk, P.G.~Blunden}{Nucl. Phys.}{A778}{44-52}{2006}
  {Calculation of two-photon exchange effects for Delta production in electron-proton collisions}
\bibitem{TBE}
  \Jou{J.A. Tjon, P.G. Blunden, W. Melnitchouk}{Phys. Rev. C}{79}{055201}{2009}
  {Detailed analysis of two-boson exchange in parity-violating e-p scattering}
\bibitem{TPE-DIS}
  \Jou{A. Afanasev, M. Strikman, C. Weiss}{Phys.Rev. D}{77}{014028}{2008}
  {Transverse target spin asymmetry in inclusive DIS with two-photon exchange}

\bibitem{Tomalak-mup}
  \Jou{O.~Tomalak, M.~Vanderhaeghen}{Phys. Rev. D}{90}{013006}{2014}
  {Two-photon exchange corrections in elastic muon-proton scattering}
\bibitem{AfanasevTri}
  \Jou{O.~Koshchii, A.~Afanasev}{Phys. Rev. D}{94}{116007}{2016}
  {Contribution of $\sigma$-meson exchange to elastic lepton-proton scattering}
\bibitem{ourTri}
  \Jou{D.~Borisyuk}{Phys. Rev. C}{96}{055201}{2017}
  {Meson exchange in lepton-nucleon scattering and the proton radius puzzle}

\bibitem{Tomalak-disp-subtracted}
  \Jou{O. Tomalak, M. Vanderhaeghen}{Eur. Phys. J. A}{51}{24}{2015}
  {Subtracted dispersion relation formalism for the two-photon exchange correction to elastic electron-proton scattering: comparison with data}
\bibitem{Tomalak-disp-muon}
  \Jou{O. Tomalak, M. Vanderhaeghen}{Eur. Phys. J. C}{78}{514}{2018}
  {Dispersion relation formalism for the two-photon exchange correction to elastic muon–proton scattering: elastic intermediate state}

\bibitem{ZhouTri}
  \Jou{H.-Y. Chen, H.-Q. Zhou}{Phys. Rev. C}{90}{045205}{2014}{}
\bibitem{Chin-deut}
  \Jou{Yu Bing Dong, D.Y. Chen}{Phys. Lett. B}{675}{426-432}{2009}
  {Two-photon exchange effect on deuteron electromagnetic form factors}
\bibitem{Chin-deut-2}
  \Jou{Yu Bing Dong}{Phys. Rev. C}{80}{025208}{2009}
  {Estimate of the two-photon exchange effect on deuteron electromagnetic form factors}
\bibitem{Kob-deut}
  \Jou{A.P. Kobushkin, Ya.D. Krivenko-Emetov, S. Dubnicka}{Phys. Rev. C}{81}{054001}{2010}
  {Elastic electron-deuteron scattering beyond one-photon exchange}
\bibitem{Kob-deut-2}  
  \Jou{A.P. Kobushkin, Ya.D. Krivenko-Emetov, S. Dubnicka, A.Z. Dubnickova}{Phys. Rev. C}{84}{054007}{2011}
  {Two-photon exchange and elastic scattering of longitudinally polarized electron on polarized deuteron}
\bibitem{Kob-tri}
  \Jou{A.P. Kobushkin, Ju. V. Timoshenko}{Phys. Rev. C}{88}{044002}{2013}
  {Two-photon exchange in electron-trinucleon elastic scattering}

\bibitem{bmtPi}
  \Jou{P.G. Blunden, W. Melnitchouk, J.A. Tjon}{Phys. Rev. C}{81}{018202}{2010}
  {Two-photon exchange corrections to the pion form factor}
\bibitem{otherPi}
  \Jou{Yu-Bing Dong, S.D.~Wang}{Phys. Lett.}{B684}{123-126}{2010}
  {Effect of two-photon exchange on the charged pion form factor}
\bibitem{ZhouPi}
  \Jou{Hai Qing Zhou}{Phys. Lett. B}{706}{82-85}{2011}
  {Gauge Invariant Two-Photon-Exchange Contributions in $e^-\pi^+ \to e^-\pi^+$}

\end{thebibliography}
\end{document}